\documentclass[aps,prb,reprint,amsmath,amssymb,superscriptaddress,twocolumn,nofootinbib]{revtex4-2}
\usepackage{graphicx}
\usepackage{dcolumn}
\usepackage{bm}
\usepackage{physics}
\usepackage{graphicx}
\usepackage{subfigure}
\usepackage{braket}
\usepackage{amsmath}
\usepackage{commath}
\usepackage{mathtools}
\usepackage{float}
\usepackage{placeins} 
\usepackage{chemformula}
\usepackage[normalem]{ulem}
\usepackage{wasysym}
\usepackage{txfonts}
\usepackage{tikz}
\usepackage{orcidlink}
\usepackage{mathrsfs}
\usepackage{makecell}
\usepackage{dsfont}

\begin{document}
	
	\preprint{APS/123-QED}
	
	\title{Electric Field-Induced Kerr Rotation on Metallic Surfaces. }
	
	\author{Farzad Mahfouzi\orcidlink{0000-0001-9297-3886}}
	\email{Farzad.Mahfouzi@nist.gov}
	\affiliation{Physical Measurement Laboratory, National Institute of Standards and Technology, Gaithersburg, MD 20899, USA.}
	\affiliation{Department of Chemistry and Biochemistry, University of Maryland, College Park, Maryland, MD 20742, USA} 
	\author{Mark D. Stiles\orcidlink{0000-0001-8238-4156}}%
	\affiliation{Physical Measurement Laboratory, National Institute of Standards and Technology, Gaithersburg, MD 20899, USA.}	
	\author{Paul M. Haney\orcidlink{0000-0001-9390-4727}}%
	\email{Paul.Haney@nist.gov}
	\affiliation{Physical Measurement Laboratory, National Institute of Standards and Technology, Gaithersburg, MD 20899, USA.}	
	\date{\today}
	\date{\today}
	
	\begin{abstract}
		
		
		We use a combination of density functional theory calculations and optical modeling to establish that the electric field-induced Kerr rotation in metallic thin films has contributions from both non-equilibrium orbital moment accumulation (arising from the orbital Edelstein effect) and a heretofore overlooked surface Pockels effect.  The Kerr rotation associated with orbital accumulation has been studied in previous works and is largely due to the dc electric field-induced change of the electron distribution function.  In contrast, the surface Pockels effect is largely due to the dc field-induced change to the wave functions. Both of these contributions arise from the dual mirror symmetry breaking from the surface and from the dc applied field. Our calculations show that the resulting Kerr rotation is due to the dc electric field modification of the optical conductivity within a couple of nanometers from the surface, consistent with the dependence on the local mirror symmetry breaking at the surface. For thin films of Pt, our calculations show that the relative contributions of the orbital Edelstein and surface Pockels effects are comparable, and that they have different effects on Kerr rotation of $s$ and $p$ polarized light, $\theta_K^s$ and $\theta_K^p$.  The orbital Edelstein effect yields similar values of $\theta_K^s$ and $\theta_K^p$, while the surface Pockels effect leads to opposing values of $\theta_K^s$ and $\theta_K^p$.
	\end{abstract}
	
	\maketitle
	
	\section{\label{sec:sec1}Introduction}
	
	The magneto-optical Kerr effect (MOKE) has emerged as a cornerstone technique in the study of magnetic materials, offering valuable insights into their properties~\cite{OPPENEER2001_book,Qiu2000_SMOKE,Ebert1990}. It is a fundamental phenomenon extensively studied for its applications in magnetic sensing, data storage, and optoelectronic devices. This effect relies on the interaction between light and the orbital part of the magnetization. In magnetic solids, the orbital moment is typically created by the spin-orbit coupling (SOC) interacting with the spin moment, but can also arise in nonmagnetic materials due to orbital Hall effects from applied electric fields. In recent years, there has been growing interest in these electric field-induced MOKE signals in non-magnetic materials~\cite{Stamm2017, Lyalin2023, Marui2024, Choi2023}, as they offer insights into spin-orbit coupling effects and potential applications in spintronics. Nevertheless, the microscopic origin of these signals, particularly in metals, remains under active debate. 
	
	This study presents a theoretical investigation of electric field-induced Kerr rotation on metallic surfaces, focusing on platinum (Pt) as a model system. Our approach includes two components.  We first perform {\it ab-initio} calculations of the Pt film's nonlocal (two-point) optical conductivity tensor in the absence and presence of an applied in-plane dc field. We then compute light scattering amplitudes using a scattering method to solve Maxwell’s equations in media described by the resulting full nonlocal dielectric tensor. The scattering amplitudes then give the Kerr rotation for both $s$- and $p$-polarized incident light.  With this approach, we can identify different microscopic contributions to the Kerr rotation and the length scale near the surface of the Pt film over which the Kerr rotation is generated.
	
	\begin{figure}%
		{\includegraphics[scale=0.32,angle=0,trim={2.5cm 1.0cm 0.0cm 1.5cm},clip,width=0.48\textwidth]{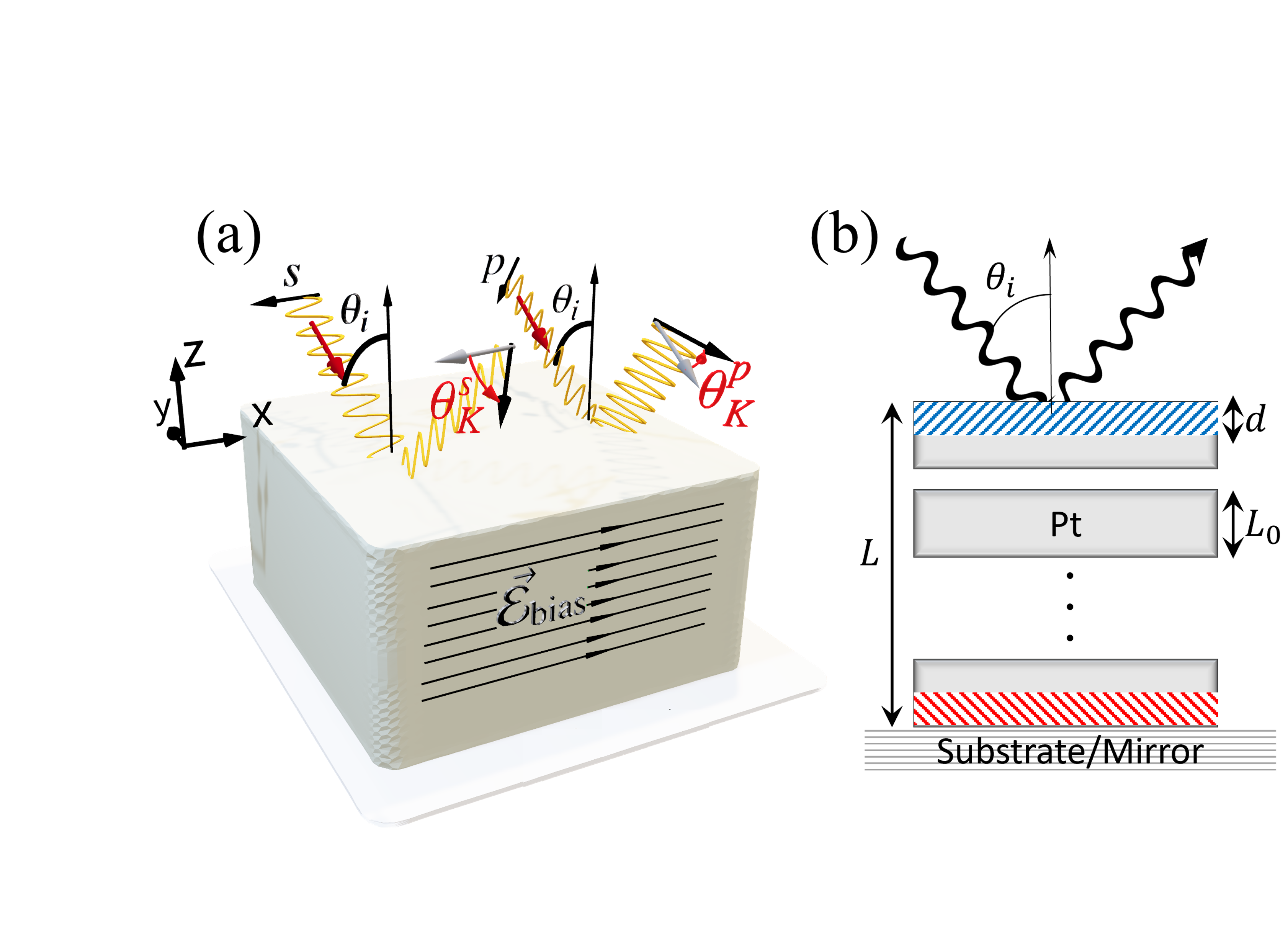}}%
		\caption{(a) Schematic depiction of the electro-optic effect. Incident light with either $s$ or $p$ polarization gets rotated in response to an in-plane bias dc electric field. (b) Schematic depiction of the calculation geometry. To calculate the Kerr rotation, we limit the electro-optic response to atomic layers near the top and bottom surface regions (with thickness $d$) as shown by the shaded areas. In order to capture optical length scales properly, we insert optically active but electro-optically inactive layers between the top and bottom films such that the total length is given by $L=N_{max}L_0$, where $N_{max}$ is the total number of repeated optical unit cells. The film is oriented with the $x$, $y$, and $z$-axes along the (100), (010), and (001) directions.}%
		\label{fig:Kerr_setup}%
	\end{figure}

	The electric field-induced change to the optical conductivity is also known as the electro-optic effect \cite{Butcher_Cotter_1990}. This effect is described with a third-rank tensor $\chi^\gamma_{\alpha\beta}(\omega)$, which relates the current oscillating at frequency $\omega$, $\vec{J}(\omega)$, to the applied dc electric field $\vec{E}_0$ and ac field, $\vec{E}(\omega)$:  \mbox{$J^\alpha(\omega) = \chi^\gamma_{\alpha\beta}(\omega) E^\beta(\omega) E^\gamma_0$}.  In this work, we keep the dc electric field spatially uniform and its direction fixed along the in-plane $x$-axis (as shown in Fig.~\ref{fig:Kerr_setup}(a)), while the ac electric field, the ac current, and the response tensor are all position dependent.  {The conventional electro-optic effect is typically considered in bulk insulators \cite{Dalton2010,Liu2015,Aillerie2000}, whereas this work focuses on the electro-optic effect in metals, localized near the sample surface. }
	
	As we describe in Sec.~\ref{sec:sec2}, our evaluation of the electro-optic effect can be framed in terms of the effect of a dc field perturbation on the ground-state ac conductivity.  We show that the static electric field modifies the ac conductivity via two distinct pathways: shifts in carrier occupancies and field-induced reshaping of the electronic wavefunctions. Within the semiclassical framework, we refer to the electro‑optic contribution arising from perturbations of the occupation function as ``extrinsic", whereas the term originating from perturbations of the wave functions is labeled as ``intrinsic". We describe the properties of each in the following paragraphs.
	
	The dc field modification of the occupation function, or extrinsic contribution to the electro-optic effect, has been studied in previous works \cite{mahfouzi2025,Manchon2024,Pesin2019,Spaldin2023}.  We show that for systems that are time-reversal invariant in the ground state, this contribution is antisymmetric in the indices of the modified ac conductivity.  By virtue of Onsager reciprocity, this implies that the extrinsic effect is related to time-reversal symmetry-breaking components of the system, which are introduced by the dc field.  This corresponds to the dc field-induced orbital magnetism, or orbital Edelstein effect \cite{mahfouzi2025}.  Interpretations of electric field-induced MOKE experiments \cite{Kato2004,Stamm2017,Lyalin2023,Marui2024} incorporate this contribution to the optical scattering, and attribute the Kerr rotation to this effect alone.

	The dc field-modification of the wave functions, or intrinsic contribution, and subsequent modification to the conductivity also plays a significant role in the Kerr rotation.  We show that this contribution leads to symmetric off-diagonal components of the dc field-modified conductivity.  This is in sharp distinction to the antisymmetric extrinsic effect; the symmetric conductivity indicates a dc modification to the system which is time-reversal symmetric.  The symmetric linear electro-optic response is the characteristic signature of the linear electro-optic (Pockels) effect.  In the Pockels effect, the breaking of the mirror plane symmetry due to an external electrostatic field in an insulator with intrinsic broken inversion symmetry results in the modification of both diagonal and symmetric off-diagonal components of the conductivity tensor.  This effect often leads to optical phenomena such as linear dichroism and birefringence. The identification of this optical response in metallic surfaces is one of the key findings of this work. 
	
	To compare with optical experiments, we incorporate the dc field-modified conductivity tensor into Maxwell's equations, numerically computing the Kerr rotation for incident $s$ and $p$ polarized light.  This calculation requires the inclusion of the nonlocal conductivity tensor and an approach for incorporating the DFT-computed conductivity tensor into simulations involving the much larger optical length scales.  With this machinery, we are able to study the extrinsic and intrinsic electro-optic contributions to the Kerr angle separately to understand the roles each plays.  We find that the extrinsic effect results in roughly equal Kerr angles for $s$ and $p$ polarized light, while the intrinsic effect yields approximately opposite Kerr angles for $s$ and $p$ polarized light. Measuring the Kerr angle for both $s$ and $p$ polarized light should allow the extrinsic and intrinsic contributions to be disentangled.  By varying the depths from the surface layer over which we include the modified conductivity tensor, we find that the Kerr angle saturates when the electro-optically active region extends a couple of nanometers from the surface, with a characteristic length that is less than 1~nm.
	
	Our findings both provide new insights into the microscopic origins of electric field-induced Kerr rotation and have broader implications for the interpretation of magneto-optical measurements in non-magnetic materials. By revealing the complex interplay between intrinsic and extrinsic contributions, this work paves the way for more accurate modeling and interpretation of magneto-optical phenomena in metallic systems.
	
	The paper is organized as follows:  In Sec.~\ref{sec:formalism1}, we present the expressions for the extrinsic and intrinsic electro-optic tensors and discuss their most important properties.  In Sec.~\ref{sec:maxwells}, we describe the approach to solving Maxwell's equations with nonlocal conductivity for realistically sized systems.  We next present the results of our calculation.  We first show the results for the electric-field induced Kerr rotation in Sec.~\ref{sec:resultsoptical}, and make comparisons to experimental data.  We next provide an in-depth analysis of the calculated electro-optic tensor for this system in Sec.~\ref{sec:eoreuslts}.  We conclude in Sec.~\ref{sec:conclusion}.  
	
	This work includes several Appendices, which we catalogue here:  App.~\ref{sec.AppA} provides some details in computational methodology related to the density functional theory calculations.  App.~\ref{sec.App_Derivation} provides the derivation the electro-optic tensors.  App.~\ref{sec:previous} relates our approach to previous work on nonlinear optical conductivity.  App.~\ref{App:AppC} derives the symmetry properties of the intrinsic and extrinsic electro-optic tensors.  App.~\ref{app:Maxwell} gives details of solving the nonlocal Maxwell's equations.  App.~\ref{app:bulkresponse} describes the origin of a bulk electro-optic spatial dispersion using a Drude model.  App.~\ref{app:equilibriumFM} shows results for the MOKE response of equilibrium ferromagnets using our numerical approach.  Finally, App.~\ref{app:finite_size_extra} shows a scheme for extrapolating the optical calculations to thicker films, which is an alternative to the scheme presented in the main text.

	\section{\label{sec:sec2}Theoretical Formalism}
	
	In this section, we first derive the two-point, frequency-dependent conductivity in equilibrium by calculating the photonic linewidth due to electron-photon interaction. The result is given by the standard expression for conductivity \cite{Kubo1957,Greenwood1958} with the addition of site-projection operators on the perturbation and response matrix elements. Next, we generalize to non-equilibrium systems by including the change in the density matrix due to the applied electric field, including both interband transitions (leading to our intrinsic effect) and intraband transitions (leading to our extrinsic effect).

	The total Hamiltonian of a periodic electronic system interacting with light is given by,
	\begin{eqnarray}
		{\mathbf{H}}_{\rm tot}(t)=\sum_{IJ{\mu\nu}}{\mathbf{c}}^{\dagger}_{I{\mu}}(t){\mathbf{H}}_{I{\mu},J{\nu}}(t){\mathbf{c}}_{J{\nu}}(t)
		+\sum_{\alpha\vec{q}}\hbar\omega_{\vec{q}}\mathbf{a}^{\dagger}_{\alpha\vec{q}}(t)\mathbf{a}_{\alpha\vec{q}}(t),
	\end{eqnarray}
	where ${\mathbf{c}}^{\dagger}_{I{\mu}}(t)$ and ${\mathbf{c}}_{I{\mu}}(t)$ are the creation and annihilation operators for an electron at time $t$, atom $I$ and atomic orbital (including spin), ${\mu}$. Here, bold symbols represent operators in Fock (many-particle) space. The photonic dispersion is denoted by $\hbar\omega_{\vec{q}}$ which has dimensions of energy, and $\mathbf{a}^{\dagger}_{\alpha\vec{q}}(t)$ and $\mathbf{a}_{\alpha\vec{q}}(t)$ are the creation and annihilation operators for a photon with polarization $\alpha$, and momentum $\vec{q}$ at time $t$. In real space, and assuming the vector potential varies slowly in space, the time-dependent potential $\vec{\mathbf{A}}_I(t)$ yields the following tight-binding Hamiltonian matrix element between orbitals $\mu$ on atom $I$ and $\nu$ on atom $J$ in a nonorthogonal atomic-orbital basis with overlap $\mathcal{S}_{I\mu,J\nu}$,
	\begin{align}
		{\mathbf{H}}_{I{\mu},J{\nu}}(t)=&{H}_{I{\mu},J{\nu}}^0+\frac{e}{2}\left[\vec{\mathbf{A}}_{I}(t)+\vec{\mathbf{A}}_J(t)\right]\cdot\vec{{v}}_{I{\mu},J{\nu}}\nonumber\\
		&+\frac{e^2}{4m}\left[\vec{\mathbf{A}}_{I}^2(t)+\vec{\mathbf{A}}_{J}^2(t)\right]{\mathcal{S}}_{I{\mu},J{\nu}},
	\end{align}
	where ${H}^0_{I{\mu}, J{\nu}}(t)$ is the electronic Hamiltonian in the absence of interactions with the photons, $\vec{{v}}_{I{\mu},J{\nu}}$ is the electronic velocity operator given in Fourier space by Eq.~\eqref{Eq.A1} in Appendix~\ref{sec.AppA}, $-e$ is the electron charge and $m$ is the electron mass. The electromagnetic vector potential at site $I$ in Fourier space is given by,
	\begin{align}
		\vec{\mathbf{A}}_I(\omega)=&\sqrt{\frac{\hbar}{2V\omega\epsilon_0}}\sum_{\alpha}\left(\mathbf{a}^{\dagger}_{\alpha I}(\omega)\vec{\rm e}^{\mbox{*}}_{\alpha}+\mathbf{a}_{\alpha I}(\omega)\vec{\rm e}_{\alpha}\right),
	\end{align}
	
	where $\epsilon_0$ is the vacuum permittivity, $V$ is the volume, and $\vec{\rm e}_{\alpha}$ is the photon polarization vector, which is complex in general.  
	
	The nonlocal optical conductivity is given by \cite{mahfouzi2025}
	\begin{flalign}\label{Eq.eq_opt_cond}
		\sigma_{I\alpha J\beta}(\omega)={\frac{\hbar e^2}{iV}}\sum_{mn}&\frac{{\hat{v}}^{I\alpha}_{mn}{\hat{v}}^{J\beta}_{nm}f_{mn}}{\varepsilon_{mn}(\hbar\omega-\varepsilon_{mn}-2i\eta)}.
	\end{flalign}
	Here, a hatted symbol denotes an operator acting in the single‑electron Hilbert space, and we define \mbox{$f_{mn}=f(\varepsilon_{m\vec{k}})-f(\varepsilon_{n\vec{k}})$} and \mbox{$\varepsilon_{mn}=\varepsilon_{m\vec{k}}-\varepsilon_{n\vec{k}}$}, with $f(E)$ being the Fermi distribution function. Since the system studied here consists of thin films that are periodic only within the plane, the Brillouin-zone sampling is restricted to in-plane wave vectors, $\vec{k}_{||}$.  Moreover, because the incident photons carry negligible in-plane momentum, $\vec{q}_{||}$, all optical matrix elements couple electronic states evaluated at the \emph{same}
	crystal momentum \(\vec{k}\). The explicit \(\vec{k}\)-index will be omitted from now on, where an averaging over \(k\)-points, $\frac{1}{N_k}\sum_{\vec{k}}$, is always understood whenever there is a summation over
	band indices \(m,n\).  The site-resolved group velocity is defined as, \mbox{$\vec{\hat{v}}^I=(\vec{\hat{v}}\hat{1}_I+\hat{1}_I\vec{\hat{v}})/2$}, where $\hat{1}_I$ is the site operator (i.e. identity matrix for orbitals corresponding to site $I$ and zero elsewhere). 
	
	\subsection{Effects of Bias Electric Field}\label{sec:formalism1}
	As shown in Appendix~\ref{sec.App_Derivation}, the change in optical conductivity in response to the external electric field, \mbox{$\chi^{\gamma}_{I\alpha J\beta}=\partial\sigma_{I\alpha, J\beta}/e\partial {E}^{\gamma}_{0}$}, can be decomposed into the following extrinsic and intrinsic components: 
	\begin{subequations}\label{Eq.neq_opt_cond}
		\begin{flalign}
			\chi^{\gamma,{\rm ext}}_{I\alpha, J\beta}&={\frac{\hbar e^2}{2iV\eta}}\sum_{mn}\frac{{\hat{v}}^{I\alpha}_{mn}{\hat{v}}^{J\beta}_{nm}}{\left(\hbar\omega-\varepsilon_{mn}-2i\eta\right)\varepsilon_{mn}}\frac{\partial f_{mn}}{\partial k_{\gamma}},\label{Eq.neq_opt_cond_a}\\
			\chi^{\gamma,{\rm int}}_{I\alpha, J\beta}&={\frac{\hbar e^2}{iV}}\sum_{mn}\frac{[\hat{O}^{\gamma},{\hat{v}}^{I\alpha}]_{mn}{\hat{v}}^{J\beta}_{nm}+{\hat{v}}^{I\alpha}_{mn}[\hat{O}^{\gamma},{\hat{v}}^{J\beta}]_{nm}}{\hbar\omega-\varepsilon_{mn}-2i\eta}\frac{f_{mn}}{\varepsilon_{mn}}\label{Eq.neq_opt_cond_b},
		\end{flalign}
	\end{subequations}
	where $[,]$ refers to the commutation relation, and we define \mbox{$\hat{O}^{\gamma}_{pq}=i\hat{v}^{\gamma}_{pq}{\rm Re}(1/(\varepsilon_{pq}+i\eta))/\varepsilon_{pq}$}. The necessity to consider the nonlocality of the off-diagonal surface electro-optic response is explained in Appendix~\ref{app:Maxwell}. The superscripts ``int" and ``ext" stand for intrinsic and extrinsic, respectively.  Throughout this manuscript, we adopt a fixed energy broadening of $\eta = 25~{\rm meV}$, which yields a room-temperature dc conductivity consistent with experimental measurements \cite{mahfouzi2025}.

	Although the full expression for the \emph{extrinsic} electro-optic response is obtained through a rigorous derivation provided in Appendices~\ref{sec.App_Derivation} and \ref{sec:previous}, it can be intuitively interpreted as the effect of a bias-driven drift current that modifies Fermi-surface electronic occupation and thereby modulates the optical absorption rate. Within the relaxation time approximation, this effect is effectively described by the change in electronic occupation, given by \mbox{$\partial f_{n\vec{k}}/e\partial{{E}}_{0}^{\gamma}=\frac{1}{2\eta}\partial f(\varepsilon_{n\vec{k}})/\partial {k}^{\gamma}$}. Consequently, the extrinsic contribution to the electro-optic effect can alternatively be obtained by substituting the Fermi distribution function in Eq.~\eqref{Eq.eq_opt_cond} with the nonequilibrium occupation, $\delta f_{n\vec{k}}$.
	
	The {\it intrinsic} electro-optic effect is also derived in Appendices~\ref{sec.App_Derivation} and \ref{sec:previous}, and can be understood intuitively as the perturbative change in electronic eigenstates and the resulting modification of the layer-projected group velocity in Eq.~\eqref{Eq.eq_opt_cond} in response to the biased electric field. This is expressed as
	\begin{flalign}\label{Eq.eq_int_opt_cond}
		\chi^{\gamma,{\rm int}}_{I\alpha, J\beta}={\frac{\hbar e^2}{iV}}\sum_{mn}&\frac{\partial\left({\hat{v}}^{I\alpha}_{mn}{\hat{v}}^{J\beta}_{nm}\right)}{e\partial {E}_{0}^{\gamma}}\frac{f_{mn}}{\varepsilon_{mn}(\hbar\omega-\varepsilon_{mn}-2i\eta)},
	\end{flalign}
	where \mbox{$\partial({{v}}^{I\alpha}_{mn})/e\partial {E}_{0}^{\gamma}=[\hat{O}^{\gamma},{\hat{v}}^{I\alpha}]_{mn}$} leads to Eq.~\eqref{Eq.neq_opt_cond_b}. Since this effect arises from the external electric field breaking the in-plane mirror symmetry, and because the out-of-plane symmetry-breaking occurs locally at the interface region of the material rather than throughout the bulk, it is referred to as the ``surface Pockels effect" \cite{PockelsBook}. This localized nature of the symmetry-breaking leads to the emergence of the electro-optic response confined to the atomically thin layers near the surface or interface region of the material. 
	
	For a system which is time-reversal invariant in its ground state, the two parts of the electro-optic tensor follow opposite exchange symmetries in their combined layer/direction indices. The intrinsic contribution is symmetric, \mbox{$\chi^{\gamma,{\rm int}}_{I\alpha,J\beta} = +\chi^{\gamma,{\rm int}}_{J\beta,I\alpha}$}, whereas the extrinsic contribution is antisymmetric,  \mbox{$\chi^{\gamma,{\rm ext}}_{I\alpha,J\beta} = -\chi^{\gamma,{\rm ext}}_{J\beta,I\alpha}$} (see Appendix~\ref{App:AppC} for a derivation). The symmetry or anti-symmetry with respect to these indices indicates time-reversal invariance or time-reversal breaking, respectively.  Therefore, the extrinsic (antisymmetric) contribution corresponds to a dc field-driven breaking of time-reversal symmetry, while the intrinsic (symmetric) contribution retains time-reversal invariance. For this reason, the extrinsic contribution can be associated with orbital accumulation (orbital Edelstein effect), which breaks time-reversal invariance, while the intrinsic contribution is associated with the Pockels effect, which retains time-reversal symmetry.
	
	The indices of our response tensor include both the direction of the applied electric field and the induced current, along with the layer position of each. We find that the position dependence includes substantial nonlocality, where an applied ac field in layer $I$ induces an ac current in layer $J\neq I$. We find that these nonlocal components make a significant contribution to the Kerr rotation, and it's therefore necessary to solve Maxwell's equations with a nonlocal dielectric tensor. In periodic systems, the real-space conductivity between sites $I,J$ is typically Fourier transformed to reciprocal space with a single wave vector $\vec q$, which corresponds to the optical wave vector.  In our case, the non-periodicity along the film normal direction requires the use of a real-space representation. In our results, we find responses that are both symmetric and antisymmetric with respect to layer indices.  The latter would correspond to odd powers of $q$ in Fourier space, which describe phenomena such as spatial dispersion \cite{Wang0223,Souza2010}. We discuss this more in Appendix~\ref{app:bulkresponse}.
	
	The necessity of a nonlocal description for the off-diagonal $xz,zx$ components of the surface electro-optic response can be understood by the fact that, at a metal–vacuum boundary, inversion symmetry breaking confines the electro–optic response to only a few surface layers \cite{Shen1989,Shen2011}. This locality causes a breakdown in the local approximation for the conductivity, $\sigma_{xz}(\vec{r})$, as seen by the following argument. At optical and lower frequencies, the electronic wavefunctions must satisfy closed boundary conditions that suppress out-of-plane directed charge flow, yielding an asymmetric off-diagonal optical conductivity, \(\sigma^{\mathrm{surf}}_{zx}\!\ll\!\sigma^{\mathrm{surf}}_{xz}\). This is in apparent contradiction to the symmetric or anti-symmetric response, for which the $xz$ and $zx$ components are equal in absolute value. (Note that intrinsic and extrinsic electro-optic responses have a different parametric dependence on $\eta$, and therefore do not generically combine in any particular way.) The resolution of this apparent contradiction lies in the breakdown of the local approximation for the conductivity, which becomes nonlocal, $\sigma_{\alpha\beta}(\vec{r},\vec{r'};\omega)$, containing terms that depend on the electromagnetic wave-vector, $q_{z}$. Time-reversal symmetry and Onsager reciprocity relate $\sigma_{xz}(\vec{r},\vec{r'})$ and  $\sigma_{zx}(\vec{r'},\vec{r})$, thus allowing the off-diagonal, $xz$ and $zx$ components at fixed $(\vec{r},\vec{r'})$ to differ substantially in amplitude without violating the fundamental time-reversal relations. 
	
	The contributions to the electro-optic effect we present here can be placed in the context of previous work on nonlinear optical conductivity, for example, found in Refs.~\cite{Sipe2000,aversa1995nonlinear,kumar2024band}.  This is presented in App.~\ref{sec:previous}, where we discuss the intraband components of $\hat{O}$, and the exclusion of terms that are linear in the broadening associated with the dc field.

	\subsection{Scattering Approach to Maxwell's Equations}\label{sec:maxwells}
	
	In this section, we describe our approach for calculating the Kerr rotation due to the applied dc field.  In addition to the inherent nonlocal nature of the surface off-diagonal electro-optic response, a primary challenge for the optical calculation is the mismatch between computationally feasible length scales for the {\it ab-initio} calculations of the conductivity --- where the thickest layers are $\leq$ 15 nm --- and the length scale for optical scattering, which is set by the skin depth and can exceed several tens of nanometers.  To bridge this gap, we employ a supercell approach where we stack atomistically-described layers to form thicker films, as shown in Fig.~\ref{fig:Kerr_setup}b. The spacing between adjacent layers is taken to be on the order of tenths of nanometers, much smaller than optical wavelengths, so that the optical field is not sensitive to the interruptions of the lattice.
	
	In the optical supercell, we employ a {\it``truncation''} scheme, where we remove the electro-optic contribution from the conductivity everywhere, except for the top half of the top layer and bottom half of the bottom layer (denoted by the blue and red hatched region, respectively, in Fig. ~\ref{fig:Kerr_setup}b).  The electro-optic effect is therefore only present at the real sample surfaces. An alternative to the truncation approach is to keep the full electro-optic tensor in every repeated unit cell and eliminate the spurious internal interfaces by a finite-size extrapolation (See Appendix~\ref{app:finite_size_extra} for details). In the optical frequency range considered here, this extrapolation yields results that are quantitatively similar to those obtained with the truncation method.

	In the absence of an external ac current and charge (with frequency $\omega$), Maxwell's equations with a nonlocal conductivity are given by:
	\begin{subequations}
		\begin{flalign}
			i\vec{\nabla} \times \vec{\mathcal{E}}(\omega;\vec{r}) &= {\omega} \vec{\mathcal{B}}(\omega;\vec{r}) \\
			i\vec{\nabla} \times \vec{\mathcal{B}}(\omega;\vec{r}) &= -\frac{\omega}{c^2}\int d\vec{r'} ~\overset\leftrightarrow{\epsilon}  (\vec{r},\vec{r'};\omega)\cdot\mathcal{\vec E}(\omega;\vec{r'})
		\end{flalign}
	\end{subequations}
	where, $\overset\leftrightarrow{\epsilon}  (\vec{r},\vec{r'};\omega)$ is the relative nonlocal dielectric tensor. { The effect of the external DC field, $\vec{E}_0$, is incorporated by modifying the nonlocal dielectric tensor to first order in the applied bias field, $\overset\leftrightarrow{\epsilon}=1+\frac{4\pi i}{\omega} \left(\overset\leftrightarrow{\sigma} + \overset\leftrightarrow{\chi^{\gamma}} {E}^{\gamma}_{0}\right)$, where $\overset\leftrightarrow{\chi^{\gamma}}$ is evaluated using Eq.~\ref{Eq.neq_opt_cond}.
	}

	Discretizing Maxwell's equations (See Appendix \ref{app:Maxwell} for details), we obtain the transfer matrix, $\hat{T}_N$,  that links the electromagnetic fields immediately above and below the $N^{\rm th}$ optical unit cell, where the first layer, $N=1$, is positioned at the top.  We denote the field below the layer with subscript $N$, which is equal to the field above the layer with subscript $N+1$. In this case, we have
	\begin{flalign}\label{Eq.0N_Waves}\begin{bmatrix}
			\mathcal{\vec E}_{N+1}/c\\
			\mathcal{\vec B}_{N+1}
		\end{bmatrix}
		=\hat{T}_N
		\begin{bmatrix}
			\mathcal{\vec E}_{N}/c\\
			\mathcal{\vec B}_{N}
		\end{bmatrix}, \ \ \ 
	\end{flalign}
	$\hat{T}_N$ is obtained by solving Maxwell's equations in the interior of an optical unit cell, which is described in Appendix~\ref{app:Maxwell}.  The $\mathcal{\vec E}$ and $\mathcal{\vec B}$ fields in vacuum are expressed in the basis of $s$ and $p$ polarized light, and $\hat{T}_N$ is constructed in laboratory coordinate, $x,y,z$, and contains the scattering (transmission and reflection) between all optical modes ({\it e.g.}, scattering from all $s,p$ incoming states to all $s,p$ outgoing states).  Within the truncation method, we include the electro-optic perturbation only where the symmetry is broken, namely in the upper half of the transfer matrix $\hat{T}_1$ for the topmost unit cell and the lower half of $\hat{T}_{N_{max}}$for the bottom unit cell. Successive multiplication of the $\hat{T}_{N}$ matrices across the film then yields the global transfer matrix, from which the overall reflection and transmission coefficients of the structure are extracted as described in Appendix.~\ref{app:Maxwell}.
	
	Once the reflection coefficients are calculated, the Kerr rotation of $s$ and $p$ polarized incident light, including their Kerr {\it angle}, $\theta^n_K$, and {\it ellipticity}, $\eta^n_K$ can be evaluated using~\cite{You1996}

	\begin{subequations}		\label{Eq:Eq9}
		\begin{flalign}
			\theta^s_K+i\eta_K^s=\arctan\left(\frac{r_{ps}}{r_{ss}}\right),\\
			\theta^p_K+i\eta^p_K=\arctan\left(\frac{r_{sp}}{r_{pp}}\right).
		\end{flalign}
	\end{subequations}
	
	\subsubsection{Simple model for optical scattering}
	
	The spatially-dependent, nonlocal dielectric function we employ in our treatment of Maxwell's equation introduces more complexity than standard treatments of optical scattering.  To provide some context and insight for how the Kerr angle depends on the conductivity, we briefly review a much simpler scattering problem, where the off-diagonal optical conductivity of the material is homogeneous and local in the film.  Specifically, consider $s$ and $p$ polarized incident light on a medium with a spatially uniform dielectric tensor, given by
	\begin{eqnarray}
		\overset\leftrightarrow{\bm{\epsilon}} = n^2 \begin{pmatrix}
			1 & 0 &  Q_{zx}^{\rm eff}\\
			0 & 1 & 0 \\
			Q_{xz}^{\rm eff} & 0 & 1 
		\end{pmatrix},
	\end{eqnarray}
where $n^2=\epsilon_{xx}$ and $Q^{\rm eff}_{\alpha\beta}=\epsilon_{\alpha\beta}/\epsilon_{xx}$.
{In the above, the diagonal components $n^2$ describe the material's inherent optical response, while the off-diagonal components are proportional to the applied dc electric field.}
The complex longitudinal Kerr rotations for $s$ and $p$ incident light scattering off a single interface between vacuum and this material are given by~\cite{OPPENEER2001_book,You1998,Hamrle_2007}:
\begin{subequations}   \label{Eq.Eq11} 
	\begin{eqnarray} 
		\tan(\theta^s_{K}+i\eta^s_{K}) &=& \frac{n}{n^2-1} ~\frac{\cos(\theta_i)\tan(\theta_t)}{\cos(\theta_t-\theta_i)}Q_{zx}^{\rm eff},   \\
		\tan(\theta^p_{K}+i\eta^p_{K}) &=& \frac{n}{1-n^2} ~\frac{\cos(\theta_i)\tan(\theta_t)}{\cos(\theta_t+\theta_i)}Q_{xz}^{\rm eff},
	\end{eqnarray}	
\end{subequations}


where $\theta_i$ is the angle of incidence and $\theta_t$ is the angle of transmission.  Near normal incidence ($\theta_i\approx 0$), the antisymmetric part of the dielectric tensor results in equal Kerr angle for $s$ and $p$ scattering, while the symmetric part leads to opposite Kerr rotation for $s$ and $p$ scattering.  In the full-wave optical simulations discussed next, we find that a roughly similar trend applies, where the antisymmetric part (extrinsic contribution) yields roughly equal Kerr angles for $s$ and $p$ incident light, while the symmetric part (intrinsic contribution) yields roughly opposite Kerr angles.  However, the oblique angle $45^\circ$ of incident light, together with the nonlocality of the dielectric function in the realistic system obscure this simple relation. In the results of the next section, we extract values of $Q^{\rm eff}_{\alpha\beta}$ from the full numerical simulations by examining the dependence of Kerr angle on $\theta_i$.

\section{Results and Discussion:}	

In this section, we quantify the dc-field-induced Kerr rotation in Pt thin films and disentangle the relative contributions of extrinsic and intrinsic mechanisms. We begin with the full results of the optical scattering, and then examine the properties of the electro-optic tensor in detail.

\begin{figure}%
	{\includegraphics[scale=0.32,angle=0,trim={1.0cm 0.0cm 0.0cm 0.0cm},clip,width=0.5\textwidth]{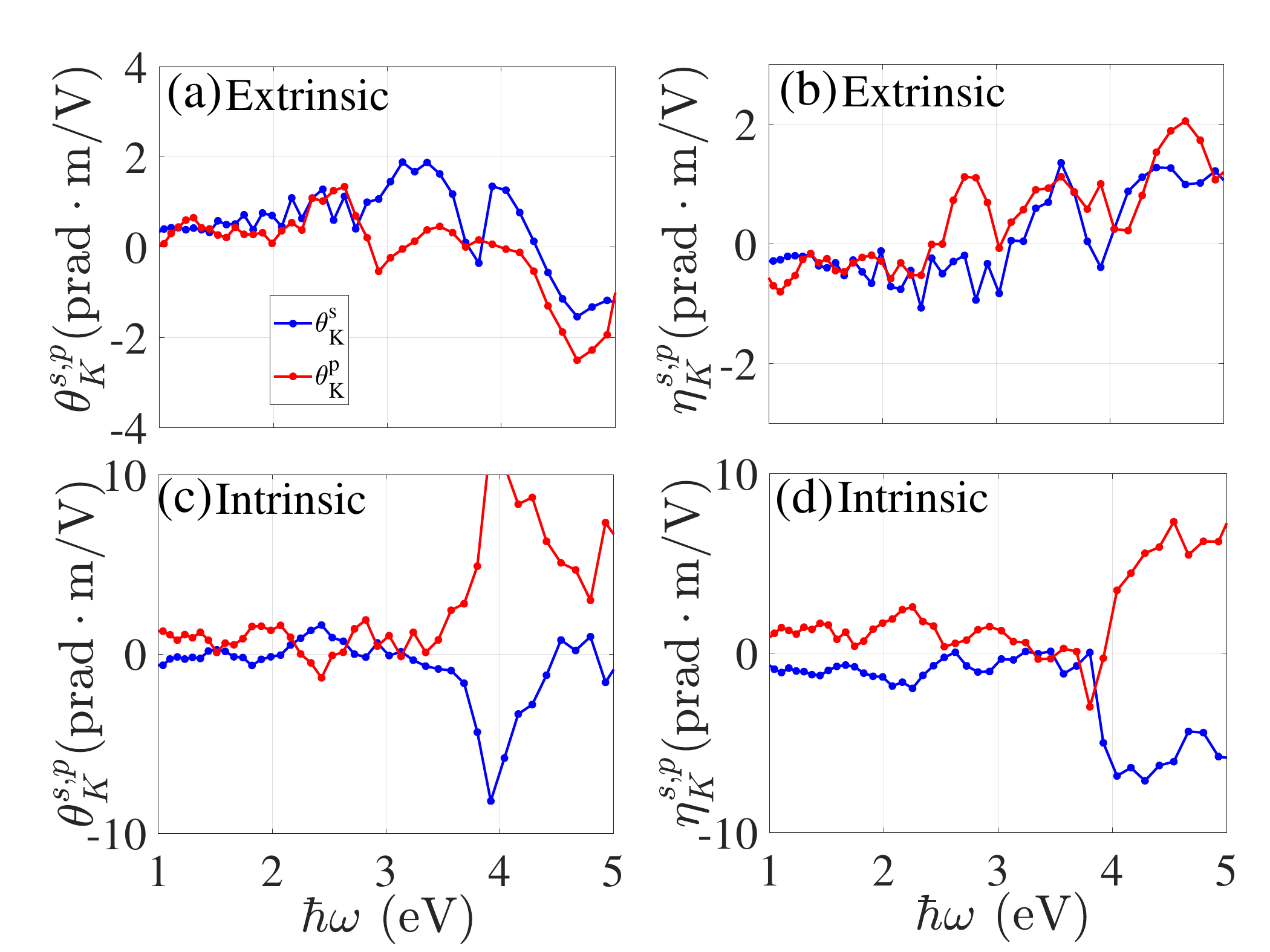}}%
	\caption{Calculated Kerr rotation due to the extrinsic and intrinsic components of the electro-optic effect in 14~nm Pt film. In (a) and (b), we show the Kerr angle and ellipticity due to extrinsic electro-optic response, while (c) and (d) present the results due to the intrinsic component. The light is incident at 45$^{\circ}$.}%
	\label{fig:fig6}%
\end{figure}

\begin{figure}%
	{\includegraphics[scale=0.32,angle=0,trim={0.0cm 0.0cm 0.0cm 0.0cm},clip,width=0.5\textwidth]{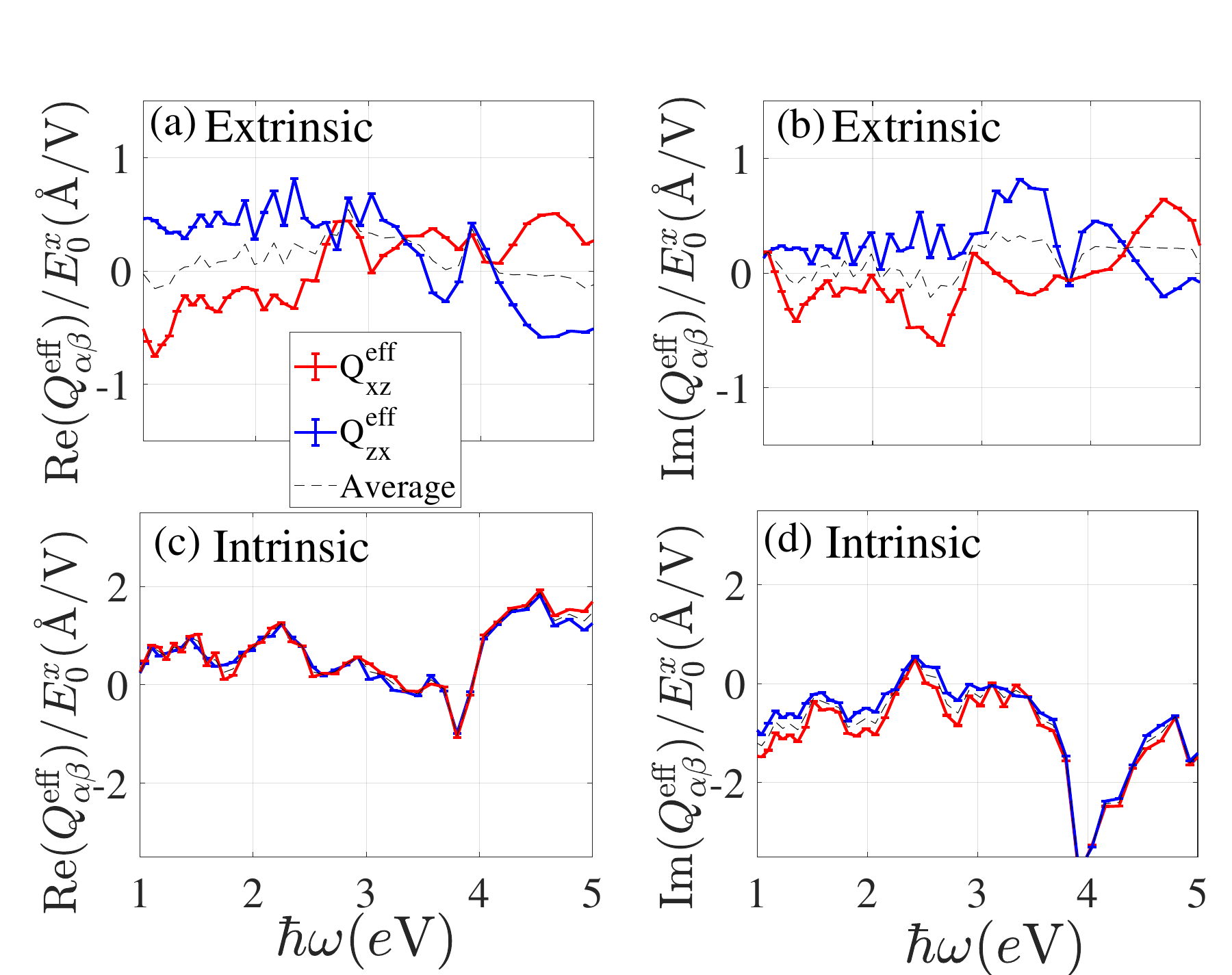}}%
	\caption{ (a) Real and (b) imaginary parts of the effective electric field induced electro-optic tensor elements, $Q^{\rm eff}_{\alpha\beta}=\epsilon_{\alpha\beta}/\epsilon_{xx}$, due to the extrinsic mechanism. The blue (red) lines show the results for $Q^{\rm eff}_{xz}$($Q^{\rm eff}_{xz}$) versus the optical frequency. The thick (thin) error-bar results correspond to 71  monolayers ($L_0=$14 nm) Pt films in the optical superlattice structure, where the error-bar amplitude is obtained from linear fitting to the angular dependence. The small error bars indicate that the results closely follow the expected angular dependence described in Eqs.~\eqref{Eq.Eq11}. Truncation method as depicted in Fig.~\ref{fig:Kerr_setup}(b) was used to extrapolate to semi-infinite Pt film thickness. (c) and (d) shows the same results due to the intrinsic mechanism.}%
	\label{fig:figQ}%
\end{figure}

\subsection{Optical results} \label{sec:resultsoptical}
Figures ~\ref{fig:fig6}(a)-(d) present the calculated dc field-induced Kerr rotation versus the optical frequency due to the intrinsic and extrinsic contributions for both $s$ and $p$ polarized light, shown as blue and red lines, respectively. We observe that the extrinsic electro-optic response results in an overall similar behavior of Kerr rotation for both $s$ and $p$ polarized incident light, in accordance with the expected Kerr rotation in systems with broken time-reversal symmetry.  On the other hand, the intrinsic electro-optic contribution produces Kerr rotations of roughly equal magnitude but opposite sign for $s$- and $p$-polarized incident light. 

Following the procedure often used in quadratic-MOKE experiments \cite{Silber2019,Hamrle_2007,Wolf2011JAP}, we first account for the different angular dependence of $s$ and $p$ polarized light to extract $Q^{\rm eff}$. To this end, the calculated complex Kerr angle,
\mbox{\(\theta^{s,p}_K+i\eta^{s,p}_K\)}, is regressed against the angle of incidence, \(\theta_i\) according to
Eqs.~\ref{Eq.Eq11}(a,b). The linear fit returns the effective field-induced electro-optic coefficients, \mbox{$Q^{\mathrm{eff}}_{\alpha\beta}=\epsilon_{\alpha\beta}/\epsilon_{xx}$}, which, by construction, are independent of \(\theta_i\).

Fig.~\ref{fig:figQ} displays the calculated \(Q^{\mathrm{eff}}_{\alpha\beta}(\omega)\) for
Pt films containing  71 
monolayers in the optical superlattice geometry.
The error-bar heights represent the 95$\%$ confidence interval obtained from the regression. The small error bars indicate that the form of Eq.~\eqref{Eq.Eq11}  describes our numerical data well.  We find that the off-diagonal elements $Q^{\rm eff}_{\alpha\beta}$, arising from the extrinsic effect, are roughly antisymmetric ({\it i.e.}, of opposite sign), as expected from the breaking of time-reversal symmetry associated with the extrinsic effect. We also find a subdominant but appreciable symmetric contribution (dotted lines in a,b) originating from the bulk.  We attribute this to significant nonlocal effects, which we discuss in more detail in the following section and in Appendix~\ref{app:bulkresponse}.  The intrinsic effect yields off-diagonal components of $Q^{\rm eff}_{\alpha\beta}$ which are almost perfectly symmetric.  This is consistent with the intrinsic contribution retaining time-reversal symmetry.

\begin{figure}
	\centering
	\includegraphics[scale=0.32,angle=0,trim={1.0cm 0.0cm 0.0cm 0.0cm},clip,width=0.45\textwidth]{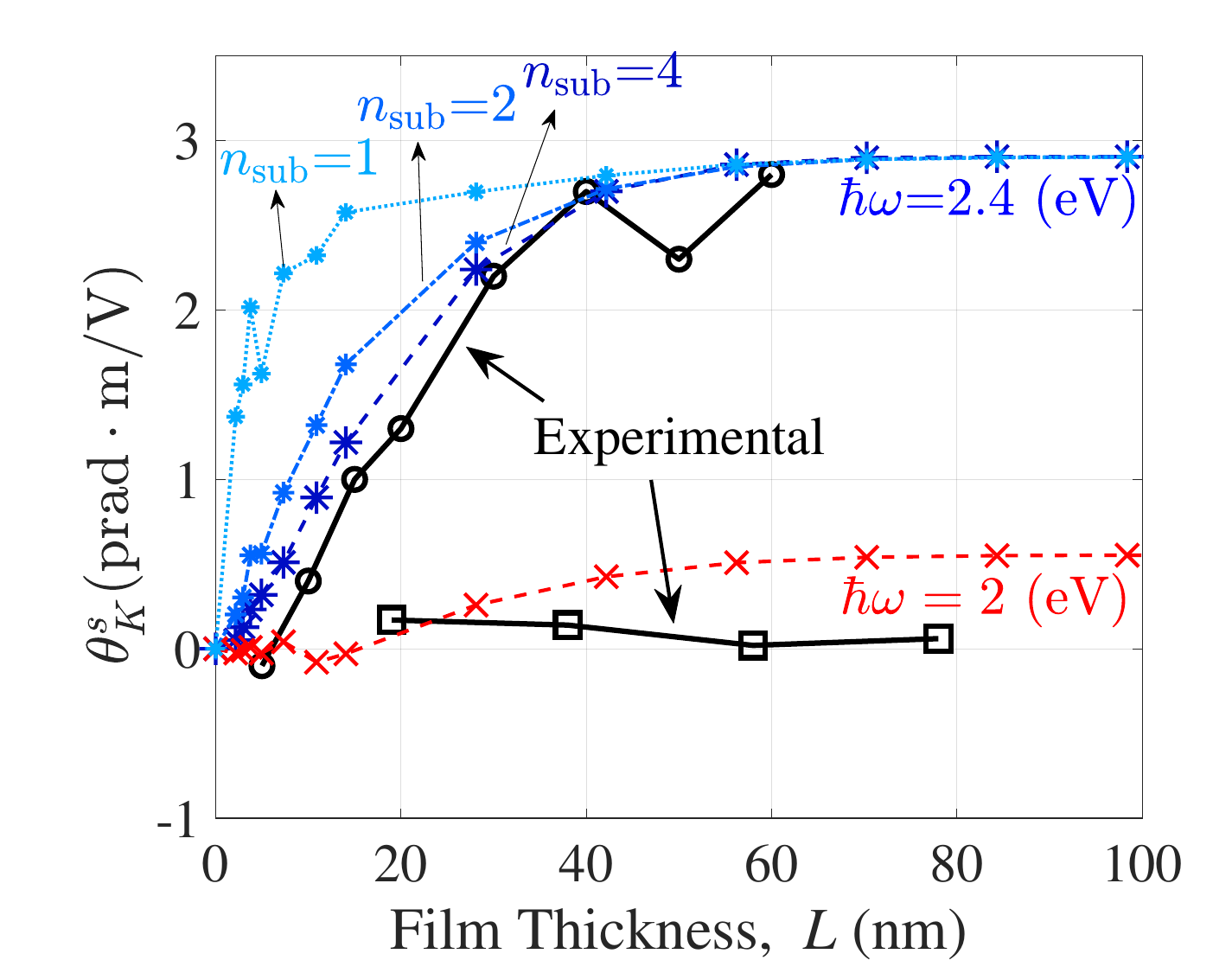}%
	\caption{Total Kerr angle for an $s$-polarized incident light versus Pt film thickness for $\hbar\omega=2$~eV and $\hbar\omega=2.4$~eV, shown as red and blue dashed lines with cross and star symbols, respectively. The thick black solid lines are from experimental measurements, as reported in Ref.~\cite{Stamm2017} and Ref.~\cite{Marui2024}. The first few points, $L<$ 20 nm films are calculated using a single Pt film, and the rest are extrapolated results obtained by constructing multiple copies of 71 monolayer Pt film stacked on each other with the truncation method for the electro-optic response as depicted in Fig.~\ref{fig:Kerr_setup}. In the case of $\hbar\omega=2.4$ eV, we consider multiple semi-infinite substrates with indices of refraction, $n_{\rm sub}=1,2,4$, corresponding to vacuum, sapphire, and silicon, respectively. An energy broadening value of $\eta=25$ meV was chosen in the numerical calculations. The incidence angle in all cases is 45$^{\circ}$.}%
	\label{fig:fig4}%
\end{figure}

\begin{figure}
	\centering
	\includegraphics[scale=0.32,angle=0,trim={1.0cm 0.0cm 0.0cm 0.0cm},clip,width=0.45\textwidth]{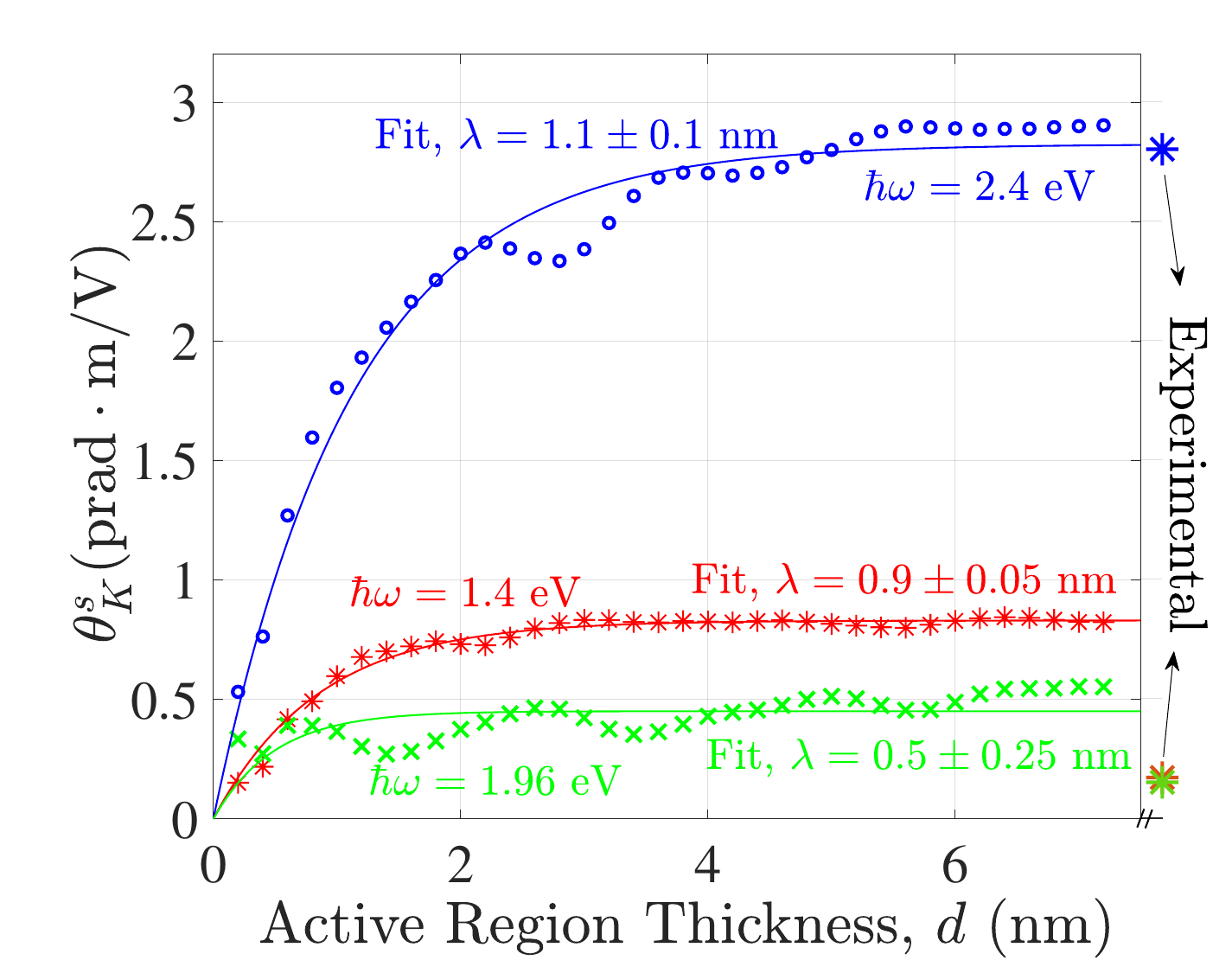}%
	\caption{Calculated Kerr angle versus thickness of the electro-optically active region near the top surface with thickness $d$, as depicted in Fig.~\ref{fig:Kerr_setup}. The solid lines correspond to the result of the fitting to the exponential decay function, $y=y_{\infty}(1-e^{-d/\lambda})$. The total thickness of the optical supercell is $\approx 0.5~{\rm \mu m}$. For reference, we also included the experimental results (star symbols), reported in Refs.~\cite{Stamm2017,Lyalin2023,Marui2024}. }
	\label{fig:fig8}%
\end{figure}
In Fig.~\ref{fig:fig4}, we compare the theoretically calculated electric field-induced Kerr angle versus Pt thickness with experimental results (shown as thick black lines) for two optical frequencies, $\hbar\omega=1.98$~eV and  $\hbar\omega=2.4$~eV, at 45$^{\circ}$ incident angle, as reported in Ref.~\cite{Marui2024} and Ref.~\cite{Stamm2017}, respectively. In the numerical calculations, the dc field-induced Kerr angles up to 71 monolayers Pt are calculated using a single film, and in order to extrapolate the results to thicker films , we constructed a super-lattice consisting of multiple layers of 71 monolayer Pt film, as described in Sec.~\ref{sec:maxwells}. We included a semi-infinite substrate with the index of refraction, $n_{\rm sub}=4$ that corresponds to Si substrate used in the experiments. For comparison, as light-blue lines in Fig.~\ref{fig:fig4}, we also present the results for the case with $n_{\rm sub}=1$ (vacuum) and  $n_{\rm sub}=2$ (sapphire).  Overall, we observe a good agreement with the reported experimental results in Refs.~\cite{Stamm2017,Marui2024}, at both probe frequencies. 

In order to investigate the effective thickness responsible for the electro-optic Kerr rotation, we change the thickness of the electro-optically active region by setting $\chi_{I\alpha,J\beta}^{\gamma}=0$ for layers $I$ beyond a distance $d$ from the surface.  The results for the Kerr angle for an $s-$polarized incident light are presented in Fig.~\ref{fig:fig8} for three probe frequencies, $\hbar\omega=$1.4~eV, 1.96~eV, and 2.4~eV. We observe an exponential dependence of the Kerr angle on $d$ which follows \mbox{$\theta_K^s=\theta_K^{s,\infty}(1-e^{-d/\lambda})$}. The solid lines in Fig.~\ref{fig:fig8} show the results of the fitting to the exponential function, where we observe a characteristic length $\lambda$ less than 1~nm, which suggests the localization of the phenomenon on the film's surface layer, consistent with the self-rotating component of the orbital Edelstein effect~\cite{mahfouzi2025}.

\subsection{Electro-optic tensor} \label{sec:eoreuslts}

\begin{figure*}[t]
	{\includegraphics[scale=1.0,angle=0,trim={2.0cm 0.0cm 1.5cm 0.0cm},clip,width=1.0\textwidth]{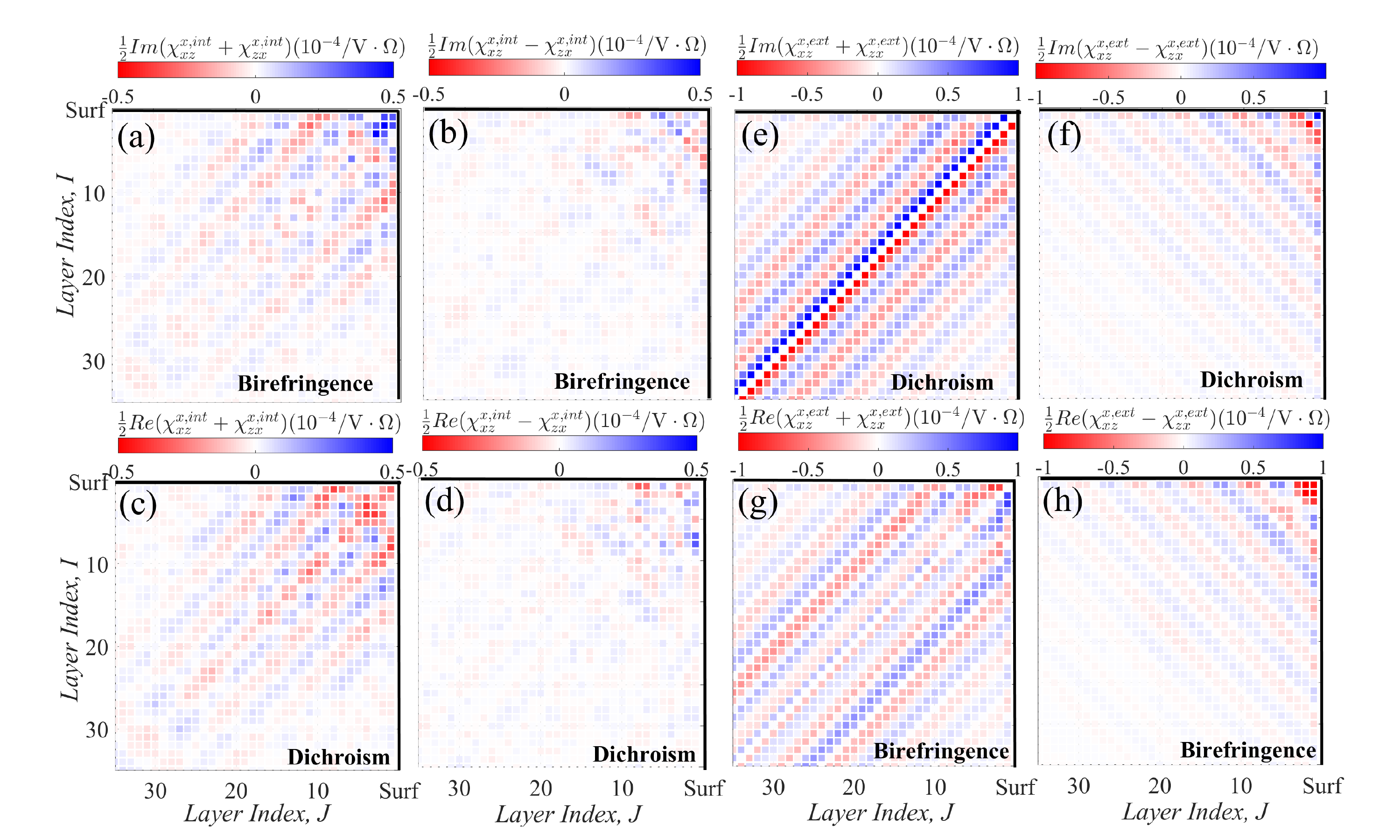}}%
	\caption{Panels (a) and (b) show the imaginary parts of the symmetric and antisymmetric intrinsic two-point electro-optic response of a 71-monolayer Pt film at photon energy $\hbar\omega$ = 2 eV; panels (c) and (d) display the corresponding real parts. Panels (e)–(h) present the same sequence for the extrinsic contribution. The results shown in panels (a,c,f,h) are layer-symmetric, while panels (b,d,e,g) depict layer-antisymmetric nonlocal optical responses. For clarity, only the upper half of the slab is plotted; the values in the lower half follow by layer inversion symmetry; responses that are layer-symmetric change sign, whereas layer-antisymmetric ones retain the same sign.}%
	\label{fig:fig9}%
\end{figure*}

We next discuss the properties of the electro-optic tensor $\chi_{I\alpha,J\beta}^{\gamma}$, computed with Eqs.~\eqref{Eq.neq_opt_cond_a} and \eqref{Eq.neq_opt_cond_b}, evaluated at \mbox{$\hbar\omega=2~{\rm eV}$}.  We take the dc field to be along the $x$-direction, and present the two-point electro-optical response function versus layer indices $I$ and $J$ in Fig.~\ref{fig:fig9}.  We show the layers extending from the surface to the middle of the film.  Panels (a-d) show the intrinsic contribution, while panels (e-h) show the extrinsic contribution.

We present the tensor in terms of components which are symmetric and antisymmetric with respect to {\it direction} indices, given by $\chi^{x}_{Ix,Jz} \pm  \chi^{x}_{Iz,Jx}$, where the + sign yields the symmetric part, and the - sign yields the antisymmetric part. Note that the resulting symmetry/anti-symmetry with respect to {\it position} indices observed in the plots are such that the intrinsic response is overall symmetric in its joint (direction, layer) indices ({\it i.e.}, even under time-reversal), and the extrinsic response is overall anti-symmetric in its joint indices ({\it i.e.}, odd under time-reversal).

We further separate the real and imaginary parts of the tensor components. Each corresponds to either dissipative or reactive parts of the ac response.  The dissipative, or dichroic part of the response is given by the Hermitian part of the electro-optic tensor: $\chi^{\gamma}_{I\alpha,J\beta}+(\chi^{\gamma}_{J\beta,I\alpha})^*$. For the extrinsic effect, this is proportional to ${\rm Im}(\chi^{\gamma,{\rm ext}}_{I\alpha,J\beta})$, while for the intrinsic effect it is proportional to ${\rm Re}(\chi^{\gamma,{\rm int}}_{I\alpha,J\beta})$.  The reactive, or birefringent, part of the response is given by the anti-Hermitian part.  In this case, the Re and Im parts are reversed. The different types of response are indicated in the plots.

The intrinsic response is shown in Figs.~\ref{fig:fig9}(a)-(d).  We find it is confined to the atomic layers near the surface.  It exhibits a larger contribution from the layer-symmetric part, shown in (a) and (c), relative to the layer-antisymmetric part, shown in (b) and (d).  For the extrinsic response, shown in Figs.~\ref{fig:fig9}(e)-(h), we observe somewhat different behavior.  The direction-symmetric components, shown in (e) and (g), exhibit layer-antisymmetric contributions that extend into the bulk of the film.  We discuss the origin of this response in Appendix~\ref{app:bulkresponse}.  Deep into the film interior, the total from this layer-antisymmetric vanishes due to compensating nonlocal currents.  However, near the surface there is a lack of cancellation due to the spatial asymmetry, and this component makes a non-negligible contribution to the Kerr rotation.

\section{\label{sec:sec4}Conclusions}	\label{sec:conclusion}
In summary, we used {\it ab-initio} methods to calculate the dc field-induced Kerr rotation in Pt films. We showed that due to the broken mirror symmetry at the metallic surfaces, the electro-optic effect results in both symmetric and antisymmetric off-diagonal contributions to the dielectric tensor, which we refer to as extrinsic and intrinsic components, respectively. While the extrinsic electro-optic tensor breaks the time-reversal symmetry, the intrinsic component does not and shares its origin with the Pockels effect. We used the {\it ab-initio} method to calculate the nonlocal electro-optic tensor and showed that intrinsic and extrinsic components have similar magnitudes. We then employed the scattering method to numerically solve Maxwell's equations in a medium with a nonlocal dielectric constant, obtaining the Kerr rotation using numerically calculated reflection coefficients. We show that for room temperature Pt films, both intrinsic and extrinsic components of the electro-optic tensor contribute to the Kerr rotation with a total Kerr angle that agrees well with the experimentally reported values. Moreover, we showed that the effective surface thickness for the optoelectronic Kerr rotation to occur is in the order of the electronic mean free path.  Our results, therefore, highlight the limitations of simplified pictures based solely on atomic-orbital Edelstein or spin/orbital Hall mechanisms; an accurate description requires the full, nonlocal electro-optical response subject to realistic surface boundary conditions. Although the present quantitative agreement with experimentally reported data for Pt is encouraging, a decisive validation will require systematic Kerr measurements on a wider range of metals and optical probe frequencies under identical growth and optical conditions.

\section{Acknowledgement}
We thank Kyung-Jin Lee and Fei Xue for fruitful discussions. We also thank Jared Wahlstrand  and
Zachary Levine for carefully reading the manuscript and providing insightful comments. FM acknowledges support under the Cooperative Research Agreement between the University of Maryland and the National Institute of Standards and Technology Physical Measurement Laboratory, Award 70NANB23H024, through the University of Maryland.

\appendix

\section{Computational Methodology}\label{sec.AppA}
The Hamiltonian $\hat{H}_{\vec{R}}$ and overlap $\hat{\mathcal{S}}_{\vec{R}}$, matrix elements of the unrelaxed (001)-oriented Pt slab { (constructed from an face-center-cubic unit cell with a lattice constant of 0.392~nm and a 1.4~nm vacuum layer)} were obtained from density functional theory calculations performed using the OpenMX  {\it ab-initio} package \cite{OzakiPRB2003,OzakiPRB2004,OzakiPRB2005}.
We adopted Troullier-Martins type norm-conserving pseudopotentials \cite{TroullierPRB1991} with partial core correction. We used a $14\times 14\times 1$ $k$-point mesh for the first Brillouin zone (BZ) integration and an energy cutoff of 500 Ry (1~Ry$\approx$13.6~eV) for numerical integrations in the real space grid. The localized orbitals were generated with radial cutoffs of 7.0~$a_0$ (1~$a_0\approx$~0.529~nm) for Pt \cite{OzakiPRB2003, OzakiPRB2004}.
We used the local spin density approximation (LSDA) \cite{CeperleyPRL1980} exchange-correlation functional as 
parameterized by Perdew and Zunger \cite{PerdewPRB1981}. 
In calculating orbital magnetization for bulk Fe, presented in Fig.~\ref{fig:fig2}(a), we used quadruple-zeta basis sets to get relative convergence with respect to the number of atomic orbitals within the linear combination of atomic orbitals (LCAO) methodology. 

The velocity operator in Fourier space was calculated using \cite{Lee2018}
\begin{flalign}			
	\hbar\vec{\hat{v}}_{\vec{k}}&=\frac{\partial \hat{H}_{\vec{k}}}{\partial\vec{k}}-\hat{H}_{\vec{k}}\hat{\mathcal{S}}_{\vec{k}}^{-1}\frac{\partial \hat{\mathcal{S}}_{\vec{k}}}{\partial\vec{k}}-i\left(\vec{\hat{r}}_{\vec{k}}\hat{\mathcal{S}}^{-1}_{\vec{k}}\hat{H}_{\vec{k}}-\hat{H}_{\vec{k}}\hat{\mathcal{S}}^{-1}_{\vec{k}}\vec{\hat{r}}_{\vec{k}}\right), \label{Eq.A1}  
\end{flalign}
where the $\vec{k}$-derivatives are obtained using
\begin{flalign}			
	\frac{\partial{\hat{C}}_{\vec{k}}}{\partial\vec{k}}&=
	i\sum_{\vec{R}}\vec{R}\hat{C}_{\vec{R}}e^{i\vec{k}\cdot\vec{R}}.
\end{flalign}
where $\hat{C}$ can represent either the Hamiltonian or the overlap matrix. The position operator within the unit cell is calculated using, $\vec{\hat{r}}_{\vec{k}}=\sum_{\vec{R}}\vec{\hat{r}}_{\vec{R}}e^{i\vec{k}\cdot\vec{R}}$, where, \mbox{$\vec{{r}}_{\vec{R}}^{\ I\mu,J\nu}=\vec{r}_I{\mathcal{S}}^{\ I\mu,J\nu}_{\vec{R}}+\vec{{r}}^{\ I\mu,J\nu}_{a,\vec{R}}$}. Here $\mu,\nu$ represent the atomic orbital basis sets, and $I,J$ are the atomic indices inside the unit cell. The position operator within the atoms is evaluated using
\begin{flalign}
	\vec{{r}}^{\ I\mu,J\nu}_{a,\vec{R}}=\int d\vec{r}\phi_{I\mu}(\vec{r})\vec{r}\phi_{J\nu}(\vec{r}-\vec{r}_{J}-\vec{R}+\vec{r}_{I}),
\end{flalign}
with, $\phi_{I\mu}(\vec{r})$, the atomic orbital basis functions for atom $I$. The electro-optic calculations were carried out using $90\times90\times1$ $k$-points mesh that we checked to be  sufficiently large enough for $\eta= 25$ meV and $\hbar\omega>1$ eV in Pt.


\section{Effects of Current Bias: Greens Function Approach}\label{sec.App_Derivation}
This section derives the expressions for the dissipative part of the optoelectronic tensor given by Eq.~\eqref{Eq.neq_opt_cond_a} and Eq.~\eqref{Eq.neq_opt_cond_b}. The photonic linewidth ({\it i.e.,} absorption rate) is related to the dissipative (Hermitian) part of the conductivity tensor, or anti-Hermitian part of the photonic self energy due to the vacuum polarization, \mbox{$\Omega^{aH}_{\alpha\beta}=( \sigma_{\alpha\beta}+ \sigma_{\beta\alpha}^*)/2\epsilon_0$}, and in a system out of equilibrium is generally given by the electron-hole recombination rate that obeys,
\begin{widetext}	
	\begin{flalign}		\label{Eq.General_diss} 
		\Omega^{aH}_{\alpha\beta}(\omega)&={\frac{\hbar e^2}{4V\hbar\omega\epsilon_0}}\int \frac{dE}{\pi}{\rm Tr}\left[{\hat{v}}^{\alpha}\left(\hat{G}^>(E){\hat{v}}^{\beta}\hat{G}^<(E+\hbar\omega)-\hat{G}^<(E){\hat{v}}^{\beta}\hat{G}^>(E+\hbar\omega)\right)\right],
	\end{flalign}
	where $\hat{G}^<$ ($\hat{G}^>$) describes the electron (hole) density matrix. It can be shown that, unlike the total optical conductivity, the diamagnetic term does not contribute to the photonic linewidth. As a result, unlike the standard velocity gauge approach \cite{Moor2019}, it remains well-behaved in the dc limit and free from divergence \cite{mahfouzi2025}. In the presence of a biased electric field, $\vec{E}_{0}$, written in the length gauge, the lesser and greater Greens functions to the lowest order are given by,
	\begin{subequations}
		\begin{flalign}		
			\hat{G}^<(E)&=2i\eta\hat{G}^rf(E-e\vec{E}_{0}\cdot\hat{\vec{r}})\hat{G}^a\nonumber\\
			&=2i{\rm Im}\left(\hat{G}^r_0\right)f(E)+2i{\rm Im}\left(\hat{G}^r_0(e\vec{E}_{0}\cdot\hat{\vec{r}})\hat{G}^r_0\right)f(E)-2i\eta\hat{G}^r_0(e\vec{E}_{0}\cdot\hat{\vec{r}})\hat{G}^a_0\frac{\partial f(E)}{\partial E},\\
			\hat{G}^>(E)&=2i{\rm Im}\left(\hat{G}^r(E)\right)-\hat{G}^<(E)\nonumber\\
			&=2i{\rm Im}\left(\hat{G}^r_0\right)(1-f(E))+2i{\rm Im}\left(\hat{G}^r_0(e\vec{E}_{0}\cdot\hat{\vec{r}})\hat{G}^r_0\right)(1-f(E))+2i\eta\hat{G}^r_0(e\vec{E}_{0}\cdot\hat{\vec{r}})\hat{G}^a_0\frac{\partial f(E)}{\partial E},
		\end{flalign}
	\end{subequations}    
	where, $\hat{G}^{r}_0=(\hat{G}^{a}_0)^{\dagger}=(E-\hat{H}-i\eta)^{-1}$ denotes the unperturbed retarded/advanced Greens function. Substituting the lesser and greater Greens function in Eq.~\eqref{Eq.General_diss}, yields the following bias-induced change in the anti-Hermitian part of the dielectric tensor,   
	\begin{flalign}
		\frac{\delta\Omega^{aH}_{\alpha\beta}(\omega)}{e{E}_{0}^{\gamma}}&={\frac{\hbar e^2}{V\hbar\omega\epsilon_0}}\int \frac{dE}{\pi}(f(E+\hbar\omega)-f(E)){\rm Tr}\left[{\hat{v}}^{\alpha}\left({\rm Im}\left(\hat{G}_0^r(E)\hat{r}^{\gamma}\hat{G}^r_0(E)\right){\hat{v}}^{\beta}{\rm Im}\left(\hat{G}_0^r(E+\hbar\omega)\right)\right)\right]\\
		&+{\frac{\hbar e^2}{V\hbar\omega\epsilon_0}}\int \frac{dE}{\pi}(f(E+\hbar\omega)-f(E)){\rm Tr}\left[{\hat{v}}^{\alpha}\left({\rm Im}\left(\hat{G}^r_0(E)\right){\hat{v}}^{\beta}{\rm Im}\left(\hat{G}^r_0(E+\hbar\omega)\hat{r}^{\gamma}\hat{G}_0^r(E+\hbar\omega)\right)\right)\right]\nonumber\\
		&+{\frac{\hbar e^2\eta}{V\hbar\omega\epsilon_0}} \int \frac{dE}{\pi}{\rm Tr}\left[{\hat{v}}^{\alpha}\left({\rm Im}\left(\hat{G}^r_0(E)\right){\hat{v}}^{\beta}\hat{G}^r_0(E+\hbar\omega)\hat{r}^{\gamma}\hat{G}^a_0(E+\hbar\omega) f'(E+\hbar\omega)-\hat{G}^r_0(E)\hat{r}^{\gamma}\hat{G}^a_0(E){\hat{v}}^{\beta}{\rm Im}\left(\hat{G}^r_0(E+\hbar\omega)\right) f'(E)\right)\right]\nonumber.
	\end{flalign}
	Using Bloch states as the basis set, we can diagonalize the Hamiltonian and carry out the integration over energy analytically to obtain
	\begin{flalign}		
		\frac{\delta\Omega^{aH}_{\alpha\beta}(\omega)}{e{E}_{0}^{\gamma}}&={\frac{\hbar e^2\pi}{VN_k\hbar\omega\epsilon_0}}\sum_{mnp\vec{k}}(f(\varepsilon_{n\vec{k}})-f(\varepsilon_{n\vec{k}}-\hbar\omega)){\hat{v}}^{\alpha}_{nm\vec{k}}{\hat{v}}^{\beta}_{pn\vec{k}}\hat{r}^{\gamma}_{mp}{\rm Re}\left(\frac{\delta(\varepsilon_{n\vec{k}}-\varepsilon_{m\vec{k}}-\hbar\omega)}{\varepsilon_{n\vec{k}}-\varepsilon_{p\vec{k}}-\hbar\omega-i\eta}+\frac{\delta(\varepsilon_{n\vec{k}}-\varepsilon_{p\vec{k}}-\hbar\omega)}{\varepsilon_{n\vec{k}}-\varepsilon_{m\vec{k}}-\hbar\omega-i\eta}\right)\nonumber\\
		&+{\frac{\hbar e^2\pi}{VN_k\hbar\omega\epsilon_0}}\sum_{mnp\vec{k}}(f(\varepsilon_{n\vec{k}}+\hbar\omega)-f(\varepsilon_{n\vec{k}})){\hat{v}}^{\alpha}_{mn\vec{k}}{\hat{v}}^{\beta}_{np\vec{k}}\hat{r}^{\gamma}_{pm}{\rm Re}\left(\frac{\delta(\varepsilon_{n\vec{k}}-\varepsilon_{m\vec{k}}+\hbar\omega)}{\varepsilon_{n\vec{k}}-\varepsilon_{p\vec{k}}+\hbar\omega-i\eta}+\frac{\delta(\varepsilon_{n\vec{k}}-\varepsilon_{p\vec{k}}+\hbar\omega)}{\varepsilon_{n\vec{k}}-\varepsilon_{m\vec{k}}+\hbar\omega-i\eta}\right)\nonumber\\
		&+{\frac{\hbar e^2\pi}{2\eta VN_k\hbar\omega\epsilon_0}}\sum_{mn\vec{k}}{\hat{v}}^{\alpha}_{mn\vec{k}}{\hat{v}}^{\beta}_{nm\vec{k}}\left(\hat{v}^{\gamma}_{mm} f'(\varepsilon_{m\vec{k}})-\hat{v}^{\gamma}_{nn} f'(\varepsilon_{n\vec{k}})\right)\delta(\varepsilon_{n\vec{k}}-\varepsilon_{m\vec{k}}+\hbar\omega)\nonumber.
	\end{flalign}
	Here, $\delta(x)={\rm Im}(1/(x-2i\eta))/\pi$ is the Dirac delta-function and we used the relations
	\begin{flalign}		
		\int_{-\infty}^{\infty} \frac{dx}{\pi} {\rm Im}(\frac{1}{x-i\eta}){\rm Im}(\frac{1}{x-a-i\eta})&={\rm Im}(\frac{1}{a-2i\eta})\\
		\int_{-\infty}^{\infty} \frac{dx}{\pi} {\rm Im}(\frac{1}{x-i\eta}){\rm Im}(\frac{1}{x-a-i\eta})^2&\approx\frac{1}{2\eta}{\rm Im}(\frac{1}{a-4i\eta/3})\approx\frac{\pi}{2\eta}\delta(a),\ \ \eta\rightarrow 0.
	\end{flalign}	
	
	Introducing the operator, $\hat{O}^{\gamma}_{mn\vec{k}}=\hat{r}^{\gamma}_{mn\vec{k}}{\rm Re}(1/(\varepsilon_{m\vec{k}}-\varepsilon_{n\vec{k}}-i\eta))$, the expression for the anti-Hermitian dielectric tensor can be simplified as,
	\begin{flalign}		
		\frac{\delta\Omega^{aH}_{\alpha\beta}(\omega)}{e{E}_{0}^{\gamma}}&={\frac{\hbar e^2\pi}{VN_k\epsilon_0}}\sum_{mn\vec{k}}\frac{f(\varepsilon_{m\vec{k}})-f(\varepsilon_{n\vec{k}})}{\varepsilon_{m\vec{k}}-\varepsilon_{n\vec{k}}}\left({\hat{v}}^{\alpha}_{mn\vec{k}}[\hat{O}^{\gamma},{\hat{v}}^{\beta}]_{nm\vec{k}}+[\hat{O}^{\gamma},{\hat{v}}^{\alpha}]_{mn\vec{k}}{\hat{v}}^{\beta}_{nm\vec{k}}\right)\delta(\varepsilon_{n\vec{k}}-\varepsilon_{m\vec{k}}+\hbar\omega)\nonumber\\
		&+{\frac{\hbar e^2\pi}{2\eta VN_k\epsilon_0}}\sum_{mn\vec{k}}{\hat{v}}^{\alpha}_{mn\vec{k}}{\hat{v}}^{\beta}_{nm\vec{k}}\frac{\hat{v}^{\gamma}_{mm} f'(\varepsilon_{m\vec{k}})-\hat{v}^{\gamma}_{nn} f'(\varepsilon_{n\vec{k}})}{\varepsilon_{m\vec{k}}-\varepsilon_{n\vec{k}}}\delta(\varepsilon_{n\vec{k}}-\varepsilon_{m\vec{k}}+\hbar\omega).
	\end{flalign}

	The Kramers-Kronig relation can then be used to obtain the reactive part of the optical self-energy, resulting in the following expression for the total (retarded) photonic self-energy,
	\begin{flalign}		
		\frac{\delta\Omega^r_{\alpha\beta}(\omega)}{e{E}_{0}^{\gamma}}&=\frac{\hbar e^2}{iVN_k\epsilon_0}\sum_{mn\vec{k}}\frac{{\hat{v}}^{\alpha}_{mn\vec{k}}[\hat{O}^{\gamma},{\hat{v}}^{\beta}]_{nm\vec{k}}+[\hat{O}^{\gamma},{\hat{v}}^{\alpha}]_{mn\vec{k}}{\hat{v}}^{\beta}_{nm\vec{k}}}{\varepsilon_{n\vec{k}}-\varepsilon_{m\vec{k}}+\hbar\omega-2i\eta}\frac{f(\varepsilon_{m\vec{k}})-f(\varepsilon_{n\vec{k}})}{\varepsilon_{m\vec{k}}-\varepsilon_{n\vec{k}}}\nonumber\\
		&+{\frac{\hbar e^2}{2i\eta VN_k\epsilon_0}}\sum_{mn\vec{k}}\frac{{\hat{v}}^{\alpha}_{mn\vec{k}}{\hat{v}}^{\beta}_{nm\vec{k}}}{\varepsilon_{n\vec{k}}-\varepsilon_{m\vec{k}}+\hbar\omega-2i\eta}\frac{\partial_{k^{\gamma}} f(\varepsilon_{m\vec{k}})-\partial_{k^{\gamma}}f(\varepsilon_{n\vec{k}})}{\varepsilon_{m\vec{k}}-\varepsilon_{n\vec{k}}}\label{Eq:EqB5}.
	\end{flalign}
	
	The first term is the intrinsic contribution, given by Eq.~\eqref{Eq.neq_opt_cond_a} of the main text, while the second term is the extrinsic contribution, given by Eq.~\eqref{Eq.neq_opt_cond_b} of the main text. 
	
\end{widetext}

\section{Relation to Previous Works}\label{sec:previous}

In this section, we relate our treatment of the electro-optic effect to that found in previous works \cite{Sipe2000,aversa1995nonlinear,kumar2024band}. Previous works often use the density matrix equation of motion.  In what follows, we relate this approach to the results we obtain with the Keldysh formalism presented in the previous section.  We start from the equation of motion for the density matrix, given by:
\begin{align*}
	i\hbar\frac{\partial \hat{\rho}(t)}{\partial t}&=[\hat{H}(t),\hat{\rho}(t)],
\end{align*}
where, 
\begin{align*}
	\hat{H}(t)=\hat{H}_0+\left(\hat{V}_{1}e^{i\hbar\omega_1 t}+\hat{V}_{2}e^{i\hbar\omega_2 t}+h.c.\right).
\end{align*}

Using the Bloch basis set that diagonalizes the unperturbed Hamiltonian, $\hat{H}_0$, we obtain
\begin{align*}
	(\hbar\omega_1+\hbar\omega_2-\varepsilon_{nm})\hat{\rho}^{(2)}_{nm}(\omega_1+\omega_2)&=[\hat{V}_{1},\hat{\rho}^{(1)}_2]_{nm}+[\hat{V}_{2},\hat{\rho}^{(1)}_1]_{nm},\\
	(\hbar\omega_i-\varepsilon_{nm})[\hat{\rho}^{(1)}_i]_{nm}=[\hat{V}_{i},\hat{\rho}^{(0)}]_{nm}&=[\hat{V}_{i}]_{nm}f_{mn}, \ \ i=1,2,
\end{align*}
where, $f_{nm}=f(\varepsilon_n)-f(\varepsilon_m)$ and $\varepsilon_{nm}=\varepsilon_n-\varepsilon_m$. The second order perturbation term to the density matrix is then given by,\begin{widetext}   
	\begin{align*}
		(\hbar\omega_1+\hbar\omega_2-\varepsilon_{nm})\hat{\rho}^{(2)}_{nm}(\omega_1+\omega_2)&=[\hat{V}_{1}]_{n\ell}[\hat{\rho}^{(1)}_2]_{\ell m}-[\hat{\rho}^{(1)}_2]_{n\ell}[\hat{V}_{1}]_{\ell m}+[\hat{V}_{2}]_{n\ell}[\hat{\rho}^{(1)}_1]_{\ell m}-[\hat{\rho}^{(1)}_1]_{n\ell}[\hat{V}_{2}]_{\ell m},\\
		&=\frac{[\hat{V}_{1}]_{n\ell}[\hat{V}_2]_{\ell m}}{\hbar\omega_2-\varepsilon_{\ell m}}f_{m\ell}-\frac{[\hat{V}_2]_{n\ell}[\hat{V}_{1}]_{\ell m}}{\hbar\omega_2-\varepsilon_{n\ell}}f_{\ell n}+\frac{[\hat{V}_{2}]_{n\ell}[\hat{V}_1]_{\ell m}}{\hbar\omega_1+\varepsilon_{\ell m}}f_{m\ell}-\frac{[\hat{V}_1]_{n\ell}[\hat{V}_{2}]_{\ell m}}{\hbar\omega_1-\varepsilon_{n\ell}}f_{\ell n},
	\end{align*}
	The expectation value of an observable operator, $\hat{V}_o$, can be calculated using
	\begin{align*}
		\langle\hat{V}_o\rangle^{(2)}={\rm Tr}[\hat{V}_o\hat{\rho}^{(2)}]=\frac{[\hat{V}_o]_{mn}}{\hbar\omega_1+\hbar\omega_2-\varepsilon_{nm}}\left(\frac{[\hat{V}_{1}]_{n\ell}[\hat{V}_2]_{\ell m}}{\hbar\omega_2-\varepsilon_{\ell m}}f_{m\ell}-\frac{[\hat{V}_2]_{n\ell}[\hat{V}_{1}]_{\ell m}}{\hbar\omega_2-\varepsilon_{n\ell}}f_{\ell n}+\frac{[\hat{V}_{2}]_{n\ell}[\hat{V}_1]_{\ell m}}{\hbar\omega_1+\varepsilon_{\ell m}}f_{m\ell}-\frac{[\hat{V}_1]_{n\ell}[\hat{V}_{2}]_{\ell m}}{\hbar\omega_1-\varepsilon_{n\ell}}f_{\ell n}\right),
	\end{align*}
	
	To extend this expression to metals and allow time-reversal–symmetry breaking, we adiabatically switch on the perturbation $\hat{V}_j$ at a rate $\eta_j\ll\hbar\omega_i$, implemented by the substitution, $\hbar\omega_j\rightarrow\hbar\omega_j-i\eta_i$, to obtain
	\begin{align}\label{EqAppndx0}
		\langle\hat{V}_o\rangle^{(2)}
		&=
		\frac{[\hat{V}_o]_{mn}}{\hbar\omega_1+\hbar\omega_{2}-\varepsilon_{nm}-i(\eta_{\rm 2}+\eta_{\rm 1})}
		\Bigl[
		\frac{[\hat{V}_1]_{n\ell} [\hat{V}_2]_{\ell m} f_{n\ell}}{\hbar\omega_1-\varepsilon_{n\ell}-i\eta_{\rm 1}}
		- \frac{[\hat{V}_2]_{n\ell} [\hat{V}_1]_{\ell m}\,f_{\ell m}}{\hbar\omega_1-\varepsilon_{\ell m}-i\eta_{\rm 1}}+
		\frac{[\hat{V}_2]_{n\ell}\,[\hat{V}_1]_{\ell m}\,f_{n\ell}}{\hbar\omega_{\rm 2}-\varepsilon_{n\ell}-i\eta_{\rm 2}}
		- 
		\frac{[\hat{V}_1]_{n\ell}\,[\hat{V}_2]_{\ell m}\,f_{\ell m}}{\hbar\omega_{\rm 2}-\varepsilon_{\ell m}-i\eta_{\rm 2}}
		\Bigr].
	\end{align}
	Using the partial‐fraction identity
	\[
	\frac1{(\omega-a)(\omega-b)}
	=\frac1{a-b}\Bigl(\frac1{\omega-a}-\frac1{\omega-b}\Bigr),
	\]
	we rewrite the first two pieces to give
	\begin{align*}
		\langle\hat{V}_o\rangle^{(2)}
		&=
		[\hat{V}_o]_{mn}
		\Biggl[
		\frac{[\hat{V}_1]_{n\ell}\,[\hat{V}_2]_{\ell m}\,f_{n\ell}}
		{\varepsilon_{n\ell}-\varepsilon_{nm}+\hbar\omega_{\rm 2}-i\eta_{\rm 2}}
		\Bigl(\frac1{\hbar\omega_1-\varepsilon_{n\ell}-i\eta_{\rm 1}}-\frac1{\hbar\omega_1+\hbar\omega_{\rm 2}-\varepsilon_{nm}-i(\eta_{\rm 2}+\eta_{\rm 1})}\Bigr)\\
		&-
		\frac{[\hat{V}_2]_{n\ell}\,[\hat{V}_1]_{\ell m}\,f_{\ell m}}
		{\varepsilon_{\ell m}-\varepsilon_{nm}+\hbar\omega_{\rm 2}-i\eta_{\rm 2}}
		\Bigl(\frac1{\hbar\omega_1-\varepsilon_{\ell m}-i\eta_{\rm 1}}-\frac1{\hbar\omega_1+\hbar\omega_{\rm 2}-\varepsilon_{nm}-i(\eta_{\rm 2}+\eta_{\rm 1})}\Bigr)
		\Biggr]\\
		&\quad
		+\;
		\frac{[\hat{V}_o]_{mn}}{\hbar\omega_1+\hbar\omega_{\rm 2}-\varepsilon_{nm}-i(\eta_{\rm 2}+\eta_{\rm 1})}
		\Bigl[
		\frac{[\hat{V}_2]_{n\ell}\,f_{n \ell}}{\hbar\omega_{\rm 2}-\varepsilon_{n\ell}-i\eta_{\rm 2}}\,[\hat{V}_1]_{\ell m}
		- [\hat{V}_1]_{n\ell}\,
		\frac{[\hat{V}_2]_{\ell m}f_{\ell m}}{\hbar\omega_{\rm 2}-\varepsilon_{\ell m}-i\eta_{\rm 2}}
		\Bigr].
	\end{align*}
	
	Collecting all terms proportional to \(\frac1{\hbar\omega_1+\hbar\omega_{\rm 2}-\varepsilon_{nm}-i(\eta_{\rm 2}+\eta_{\rm 1})}\):
	\begin{align*}
		\langle\hat{V}_o\rangle^{(2)}
		&=
		[\hat{V}_o]_{mn}
		\Biggl[
		\frac{[\hat{V}_1]_{n\ell}\,[\hat{V}_2]_{\ell m}\,f_{n\ell}}{\varepsilon_{m\ell}+\hbar\omega_{\rm 2}-i\eta_{\rm 2}}
		\,\frac1{\hbar\omega_1-\varepsilon_{n\ell}-i\eta_{\rm 1}}
		\;-\;
		\frac{[\hat{V}_2]_{n\ell}\,[\hat{V}_1]_{\ell m}\,f_{\ell m}}{\varepsilon_{\ell n}+\hbar\omega_{\rm 2}-i\eta_{\rm 2}}
		\,\frac1{\hbar\omega_1-\varepsilon_{\ell m}-i\eta_{\rm 1}}
		\Biggr]\\
		&\quad
		+\;
		\frac{[\hat{V}_o]_{mn}}{\hbar\omega_1+\hbar\omega_{\rm 2}-\varepsilon_{nm}-i(\eta_{\rm 2}+\eta_{\rm 1})}
		\Bigl[
		\frac{[\hat{V}_2]_{n\ell}\,[\hat{V}_1]_{\ell m}\,f_{n\ell}}{\hbar\omega_{\rm 2}+\varepsilon_{\ell n}-i\eta_{\rm 2}}-\frac{[\hat{V}_1]_{n\ell}\,[\hat{V}_2]_{\ell m}\,f_{\ell m}}{\hbar\omega_{\rm 2}+\varepsilon_{m\ell}-i\eta_{\rm 2}}
		-\;\frac{[\hat{V}_1]_{n\ell}\,[\hat{V}_2]_{\ell m}\,f_{n\ell}}{\varepsilon_{m\ell}+\hbar\omega_{\rm 2}-i\eta_{\rm 2}}
		+\frac{[\hat{V}_2]_{n\ell}\,[\hat{V}_1]_{\ell m}\,f_{\ell m}}{\varepsilon_{\ell n}+\hbar\omega_{\rm 2}-i\eta_{\rm 2}}
		\Bigr].
	\end{align*}
	Re‐index the first two terms so they carry \(f_{nm}\):
	\begin{align*}
		\langle\hat{V}_o\rangle^{(2)}
		&=
		\frac{[\hat{V}_1]_{nm} f_{nm}}{\hbar\omega_1-\varepsilon_{nm}-i\eta_{\rm 1}}
		\Bigl[
		\frac{[\hat{V}_o]_{\ell n}\,[\hat{V}_2]_{m\ell}}{\varepsilon_{\ell m}+\hbar\omega_{\rm 2}-i\eta_{\rm 2}}
		-\frac{[\hat{V}_o]_{m\ell}\,[\hat{V}_2]_{\ell n}\,}{\varepsilon_{n\ell}+\hbar\omega_{\rm 2}-i\eta_{\rm 2}}
		\Bigr]\\
		&+\;
		\frac{[\hat{V}_o]_{mn}f_{nm}}{\hbar\omega_1+\hbar\omega_{\rm 2}-\varepsilon_{nm}-i(\eta_{\rm 2}+\eta_{\rm 1})}\Bigl[
		\frac{[\hat{V}_2]_{n\ell}\,[\hat{V}_1]_{\ell m}\,}{\hbar\omega_{\rm 2}+\varepsilon_{\ell n}-i\eta_{\rm 2}}
		-\frac{[\hat{V}_1]_{n\ell}\,[\hat{V}_2]_{\ell m}\,}{\hbar\omega_{\rm 2}+\varepsilon_{m\ell}-i\eta_{\rm 2}}
		\Bigr].
	\end{align*}
	Defining
	\(\;O^{\rm tot}_{m\ell}=\frac{[\hat{V}_2]_{m\ell}}{\varepsilon_{\ell m}+\hbar\omega_{\rm 2}-i\eta_{\rm 2}}\), we obtain,  
\end{widetext}
\begin{align}\label{Eq.EqApp2}
	\langle\hat{V}_o\rangle^{(2)}
	&=\frac{[\hat{O}^{\rm tot},\,\hat{V}_o]_{mn}[\hat{V}_1]_{nm}f_{nm}}
	{\hbar\omega_1-\varepsilon_{nm}-i\eta_{\rm 1}}+\frac{[\hat{V}_o]_{mn}\,[\hat{O}^{\rm tot},\,\hat{V}_1]_{nm}f_{nm}}
	{\hbar\omega_1+\hbar\omega_{\rm 2}-\varepsilon_{nm}-i(\eta_{\rm 2}+\eta_{\rm 1})}.
\end{align}
In addition to helping in the time-reversal decomposition (as shown below), Eq.~\eqref{Eq.EqApp2} is computationally more efficient than Eq.~\eqref{EqAppndx0} because the $\ell$-sum can be precomputed once (independently of the chemical potential and the probe frequency $\omega_1$) and the chemical-potential dependence introduced only after completing the band-index matrix operations. This efficiency is particularly important for layer-resolved electro-optic calculations.

The time-reversal decomposition ${O^{\rm tot}}$ is obtained by writing \mbox{\(\;O^{\rm tot}_{m\ell}=O_{m\ell}+iO'_{m\ell}\)}, where, \mbox{\(\;O_{m\ell}=[\hat{V}_2]_{m\ell}{\rm Re}(\frac{1}{\varepsilon_{\ell m}+\hbar\omega_{\rm 2}-i\eta_{\rm 2}})\)}, is the intrinsic component (i.e., even function of $\eta_{\rm 2}$) and \mbox{\(\;O'_{m\ell}=[\hat{V}_2]_{m\ell}\pi\delta(\varepsilon_{\ell m}+\hbar\omega_{\rm 2})\)}, is the extrinsic component (i.e., odd function of $\eta_{\rm 2}$). In the limit \(\omega\equiv\omega_{1}\gg\eta_{1}/\hbar\) and \((\omega_{2},\eta_{1},\eta_{2})\to 0\), with the identifications
\[
\hat{V}_o\equiv \hat{v}^{\alpha},\qquad
\hat{V}_1\equiv \frac{i\,\hat{v}^{\beta}}{\hbar\omega},\qquad
\hat{V}_2\equiv \hat{r}^{\gamma}\equiv \frac{i\,\hat{v}^{\gamma}}{\hbar\omega_{2}},
\]
one obtains the electro-optic response given by Eqs.~\eqref{Eq.neq_opt_cond_a} and \eqref{Eq.neq_opt_cond_b}. Here we retain only the leading order in \(\eta_{2}\) for both intrinsic and extrinsic contributions, whose physical interpretations are detailed in the main text. 

If $\eta_2$ is greater than intra-band transition energies (which are equal to $\varepsilon_{n,\vec{k}}-\varepsilon_{n,\vec{k}+\vec{q}}$ where $\vec{q}$ is the optical wave vector), then the contribution from the intra-band component of $\hat{O}^{\rm int}$ is suppressed.  This is the scenario considered in this work.  On the other hand, for $\eta_2$ less than the intra-band transition energies, $\hat{O}^{\rm int}$ makes additional contributions to the response. In other words, there is a difference according to the order in which the limits $\vec{q}\to 0$ and $\eta_2\to0$ are taken.  In this work, we consider $\vec{q}\to 0$ first, and intra-band contributions do not appear.

For the extrinsic component, the delta-function \mbox{$\delta(\varepsilon_{\ell m}+\hbar\omega_{\rm 2})$} guarantees gauge equivalence between length and velocity formulations and permits the substitution, $\hat{r}^{\gamma}=i\hat{v}^{\gamma}/\hbar\omega_2$, in the derivation of Eq. (5b). There are higher-order corrections with respect to the parameter $\eta_2$, originating from the finite width of the Lorentzian in the definition of $O'_{m\ell}$ and from the expansion of the second term in Eq.~\eqref{Eq.EqApp2}.  These terms are linear in $\eta_2$.  We do not include these terms, as they are prone to violating conservation laws and gauge invariance. A detailed analysis of their effects and physical interpretations is the subject of future work.

\section{Properties of electro-optic tensors}\label{App:AppC}

In this section, we show that, for time-reversal invariant systems, the intrinsic and extrinsic electro-optic tensors are symmetric and antisymmetric in their indices, respectively.  In what follows, we write the extrinsic and intrinsic components as the product of two complex numbers, one equal to the in product of velocity matrix elements, and the other comprising the remaining factors, related to energy eigenvalues and the optical frequency. For example, the extrinsic conductivity is:
\begin{eqnarray}
	\chi^{\gamma,{\rm ext}}_{I\alpha J\beta}&=&\left[{\hat{v}}^{I\alpha}_{mn}{\hat{v}}^{J\beta}_{nm} \frac{\partial f_{mn}}{\partial k^{\gamma}}\right]\left({\frac{\hbar e^2}{2iV\eta}}\frac{1}{\left(\hbar\omega-\varepsilon_{mn}-2i\eta\right)\varepsilon_{mn}}
	\right).
	\nonumber 
\end{eqnarray}
The factor $\partial f_{mn}/\partial k^\gamma$ includes velocity matrix elements, and is given by:
\begin{eqnarray}
	\frac{\partial f_{mn}}{\partial k^{\gamma}} = v^\gamma_{mm} \frac{\partial f_m}{\partial \mu} - v^\gamma_{nn} \frac{\partial f_n}{\partial \mu}~.
\end{eqnarray}
Consider the contribution from states at ${\vec k}$ and their Kramers pairs at $-{\vec k}$. The factor in the electro-optic response that depends only on energy eigenvalues is identical for the Kramer's pair.  For the other factor, consisting of a product of velocity matrix elements, we note the following relation between a velocity matrix element between states at $\vec k$ and their Kramers pairs at $-\vec k$~\cite{Sipe2000}:
\begin{eqnarray}
	v_{nm}^{I\alpha}(\vec k) =  -v_{nm}^{I\alpha}(-\vec k)^*~.
\end{eqnarray}
Therefore, the sum over a Kramer's pair of states includes only the imaginary part of the velocity matrix element product:
\begin{eqnarray}
	\chi^{\gamma,{\rm ext}}_{I\alpha J\beta}({\vec k})+\chi^{\gamma,{\rm ext}}_{I\alpha J\beta}(-{\vec k}) \propto {\rm Im}\left[{\hat{v}}^{I\alpha}_{mn}{\hat{v}}^{J\beta}_{nm} \frac{\partial f_{mn}}{\partial k^{\gamma}}\right].
\end{eqnarray}
We will make use of this property in what follows.
\\

Next we compare $\chi^{\gamma,{\rm ext}}_{I \alpha J\beta}$ with its index-transposed counterpart $\chi^{\gamma,{\rm ext}}_{J\beta, I \alpha }$.  Evaluating Eq.~\ref{Eq.neq_opt_cond_a} for each leads to:
\begin{eqnarray}
	\chi^{\gamma,{\rm ext}}_{I \alpha, J\beta} &\propto& {\hat v}^{I\alpha}_{nm} {\hat v}^{J\beta}_{mn}\frac{\partial f_{mn}}{\partial k^{\gamma}}\\
	\chi^{\gamma,{\rm ext}}_{J\beta, I \alpha } & \propto& {\hat v}^{J\beta}_{nm} {\hat v}^{I\alpha}_{mn}\frac{\partial f_{mn}}{\partial k^{\gamma}} = \left[{\hat v}^{I\alpha}_{nm} {\hat v}^{J\beta}_{mn}\frac{\partial f_{mn}}{\partial k^{\gamma}}\right]^*\label{eq:ext2}
\end{eqnarray}
The second equality on the right-hand side of Eq.~\ref{eq:ext2} follows from the Hermitian property of the velocity operator.  By virtue of the fact that only the imaginary part of the matrix element product enters into the sum over Kramer's pairs, the above immediately implies
\begin{eqnarray}
	\chi^{\gamma,{\rm ext}}_{I \alpha, J\beta} =-\chi^{\gamma,{\rm ext}}_{J\beta, I \alpha}.\end{eqnarray}

For the intrinsic contribution, the same argument can be used.  We evaluate both $\chi^{\gamma,{\rm int}}_{I \alpha J\beta}$ with  $\chi^{\gamma,{\rm int}}_{J\beta, I \alpha }$ using Eq.~\ref{Eq.neq_opt_cond_b}.  For $\chi^{\gamma,{\rm int}}_{J\beta, I \alpha }$, we utilize the Hermitian property of the velocity operators to transpose indices of each matrix element:  $\hat{v}^{I\alpha}_{nm} \rightarrow (\hat{v}^{I\alpha}_{mn})^*$,  $\hat{v}^\gamma_{nm}=(\hat{v}^\gamma_{mn})^*$.  After some algebra, we find that the product of matrix elements for $\chi^{\gamma,{\rm int}}_{J\beta, I \alpha }$ is the negative complex conjugate of that of $\chi^{\gamma,{\rm int}}_{I \alpha ,J\beta }$.  Therefore, summing over Kramer's pairs immediately yields
\begin{eqnarray}
	\chi^{\gamma,{\rm int}}_{I \alpha, J\beta} = +\chi^{\gamma,{\rm int}}_{J\beta, I \alpha}.
\end{eqnarray}

\section{Solving Maxwell's Equations With Nonlocal Dielectric Tensor}\label{app:Maxwell}

In the absence of an external ac current and charge (with frequency $\omega$), Maxwell's equations in a linear medium are given by,

\begin{subequations}
	\begin{flalign}
		i\vec{\nabla} \times \vec{\mathcal{E}}(\omega;\vec{r}) &= {\omega} \vec{\mathcal{B}}(\omega;\vec{r}) \\
		i\vec{\nabla} \times \vec{\mathcal{B}}(\omega;\vec{r}) &= -\frac{\omega}{c^2}\sum_{\alpha\beta}\vec{e}_{\alpha}\int d\vec{r'} \epsilon^r_{\alpha\beta}(\omega;\vec{r},\vec{r'})\mathcal{E}_{\beta}(\omega;\vec{r'})
	\end{flalign}
\end{subequations}
where, $\epsilon^r_{\alpha\beta}(\omega;\vec{r},\vec{r'})$ is the relative nonlocal dielectric tensor.
The discretized  wave equation for the electric and magnetic fields on each layer along the $z$-axis, is then given by,
\begin{subequations}
	\begin{flalign}
		&i\frac{\mathcal{E}_{I+1x}-\mathcal{E}_{Ix}}{z_{I+1}-z_{I}}-q_x\frac{\mathcal{E}_{I+1z}+\mathcal{E}_{Iz}}{2}+{\omega}\frac{\mathcal{B}_{I+1y}+\mathcal{B}_{Iy}}{2}=0,\\
		&i\frac{\mathcal{E}_{I+1y}-\mathcal{E}_{Iy}}{z_{I+1}-z_{I}}-q_y\frac{\mathcal{E}_{I+1z}+\mathcal{E}_{Iz}}{2}-{\omega}\frac{\mathcal{B}_{I+1x}+\mathcal{B}_{Ix}}{2}=0,\\
		&{\omega}\frac{\mathcal{B}_{I+1z}+\mathcal{B}_{Iz}}{2}+q_x\frac{\mathcal{E}_{I+1y}+\mathcal{E}_{Iy}}{2}-q_y\frac{\mathcal{E}_{I+1x}+\mathcal{E}_{Ix}}{2}=0,\\
		&i\frac{\mathcal{B}_{I+1x}-\mathcal{B}_{Ix}}{z_{I+1}-z_{I}}-q_x\frac{\mathcal{B}_{I+1z}+\mathcal{B}_{Iz}}{2}-\frac{\omega}{c^2}\sum_{J\beta}\frac{\epsilon^r_{I+1y,J\beta}+\epsilon^r_{Iy,J\beta}}{2}\mathcal{E}_{J\beta}=0,\\
		&i\frac{\mathcal{B}_{I+1y}-\mathcal{B}_{Iy}}{z_{I+1}-z_{I}}-q_y\frac{\mathcal{B}_{I+1z}+\mathcal{B}_{Iz}}{2}+\frac{\omega}{c^2}\sum_{J\beta}\frac{\epsilon^r_{I+1x,J\beta}+\epsilon^r_{Ix,J\beta}}{2}\mathcal{E}_{J\beta}=0,\\
		&q_y\frac{\mathcal{B}_{I+1x}+\mathcal{B}_{Ix}}{2}-q_x\frac{\mathcal{B}_{I+1y}+\mathcal{B}_{Iy}}{2}+\frac{\omega}{c^2}\sum_{J\beta}\frac{\epsilon^r_{I+1z,J\beta}+\epsilon^r_{Iz,J\beta}}{2}\mathcal{E}_{J\beta}=0.
	\end{flalign}
\end{subequations}
This system of equations can be rewritten in a more compact form as follows,
\begin{flalign}\label{Eq.discMaxwell}
	\sum_{J}(\frac{\hat{\mathcal{M}}_{I+1,J}+\hat{\mathcal{M}}_{I,J}}{2}+i\hat{F}\frac{\delta_{I+1,J}-\delta_{IJ}}{z_{I+1}-z_{I}})\begin{bmatrix}
		\mathcal{\vec{E}}_{J}/c\\
		\mathcal{\vec{B}}_{J}
	\end{bmatrix}
	=0,
\end{flalign}
where, 
\begin{subequations}		
	\begin{flalign}
		\hat{F}^{\alpha\beta}&=\delta_{\alpha\beta}\hat{1},\ \ \alpha=x,y\\
		\hat{F}^{z\beta}&=\hat{F}^{\beta z}=0,\label{eq:D4b}\\
		\hat{\mathcal{M}}_{I,J}^{x\beta}&=\begin{bmatrix}
			-q_x\delta_{z\beta}\delta_{IJ} & \frac{\omega}{c}\delta_{y\beta}\delta_{IJ}\\
			-\frac{\omega}{c}\epsilon^r_{Iy,J\beta} & -q_x\delta_{z\beta}\delta_{IJ}
		\end{bmatrix},\\
		\hat{\mathcal{M}}_{I,J}^{y\beta}&=\begin{bmatrix}
			-q_y\delta_{z\beta}\delta_{IJ} & -\frac{\omega}{c}\delta_{x\beta}\delta_{IJ}\\
			\frac{\omega}{c}\epsilon^r_{Ix,J\beta} & -q_y\delta_{z\beta}\delta_{IJ}
		\end{bmatrix},\\
		\hat{\mathcal{M}}_{I,J}^{z\beta}&=\delta_{IJ}\left(q_x\delta_{y\beta}-q_y\delta_{x\beta}\right)\hat{1}\nonumber\\
		&+\frac{\omega}{c}\begin{bmatrix}
			0 & \delta_{z\beta}\delta_{IJ}\\
			-\epsilon^r_{Iz,J\beta} & 0\\
		\end{bmatrix}.
	\end{flalign}
\end{subequations}

For a homogeneous material, we have, $\hat{\mathcal{M}}^{\alpha\beta}_{I,J}=\hat{\mathcal{M}}^{\alpha\beta}\delta_{I,J}$. In this case, in Fourier space and the limit of small $q_z$, Eq.~\eqref{Eq.discMaxwell} becomes,
\begin{flalign}\label{Eq.FourierMaxwell}
	\left(\hat{\mathcal{M}}_{\vec{q}_{||}}(\omega)-\hat{F}q_z\right)\begin{bmatrix}
		\mathcal{\vec{E}}_{\vec{q}}/c\\
		\mathcal{\vec{B}}_{\vec{q}}
	\end{bmatrix}
	=0.
\end{flalign}
The resulting generalized eigenvalue problem can be solved to obtain six eigenvalues, $q_{z;n}$, and the corresponding eigenstates, $\hat{\varphi}_n$. However, since the $\hat{F}$ matrix is diagonal with only four nonzero elements, Eq.~\eqref{Eq.FourierMaxwell} is guaranteed to have at most four eigenmodes with finite eigenvalues, two of which correspond to the right-moving, $\hat{\varphi}_{1,2}^+$, and the other two describe left-moving photons, $\hat{\varphi}_{1,2}^-$. It is thus numerically advantageous to reduce the size of the matrices in the eigenvalue problem to $4\times4$ by removing, say, the z-component of the electromagnetic field,
\begin{flalign}
	\begin{bmatrix}
		\mathcal{E}^z_{\vec{q}}/c\\
		\mathcal{B}^z_{\vec{q}}
	\end{bmatrix}=-[\hat{\mathcal{M}}^{zz}_{\vec{q}_{||}}]^{-1}\sum_{\beta=x,y}\hat{\mathcal{M}}^{z\beta}_{\vec{q}_{||}}(\omega)\begin{bmatrix}
		\mathcal{E}^{\beta}_{\vec{q}}/c\\
		\mathcal{B}^{\beta}_{\vec{q}}
	\end{bmatrix}.
\end{flalign}
The effective eigenvalue problem to be solved is therefore given by,
\begin{flalign}\label{Eq.EffFourierMaxwell}
	\sum_{\beta}(\hat{\mathcal{M}}^{\alpha\beta}-\hat{\mathcal{M}}^{\alpha z}\frac{1}{\hat{\mathcal{M}}^{zz}}\hat{\mathcal{M}}^{z\beta})\begin{bmatrix}
		\mathcal{E}^{\beta}/c\\
		\mathcal{B}^{\beta}
	\end{bmatrix}
	=q_z\begin{bmatrix}
		\mathcal{E}^{\alpha}/c\\
		\mathcal{B}^{\alpha}
	\end{bmatrix}.
\end{flalign}
where, $\alpha,\beta=x,y$. In the case of incident light in the $yz$-plane in a vacuum with incident angle, $\theta_i$, for the in-plane wavevector, we have $q_x=0$, $q_y=\omega\cos(\theta_i)/c$. The out-of-plane wavevector is then obtained by solving the eigenvalue problem, Eq.~\eqref{Eq.EffFourierMaxwell}, which yields the expected four doubly degenerate eigenvalues, $q^{\pm}_z=\pm\omega\sin(\theta_i)/c$, with the corresponding eigenstates given by,
\begin{flalign}\label{Eq.EigenMaxwell}
	[\hat{\varphi}]^{\pm}_s=\begin{bmatrix}
		\mathcal{E}^x/c\\
		\mathcal{E}^y/c\\
		\mathcal{B}^x\\
		\mathcal{B}^y
	\end{bmatrix}=\begin{bmatrix}
		1\\
		0\\
		0\\
		\frac{ cq^{\pm}_z}{\omega}
	\end{bmatrix},\ \ \
	[\hat{\varphi}]^{\pm}_p=\begin{bmatrix}
		0\\
		-\frac{cq^{\pm}_z}{\omega}\\
		1\\
		0
	\end{bmatrix},
\end{flalign}

The dual of the basis set is then given by,
\begin{subequations}	
	\begin{flalign}\label{Eq.invEigenMaxwell}
		[\hat{\varphi}^{-1}]^{\pm}_s=\frac{1}{2}\begin{bmatrix}
			1,0,0,{\omega}/{cq^{\pm}_z}
		\end{bmatrix},\\
		[\hat{\varphi}^{-1}]^{\pm}_p=\frac{1}{2}\begin{bmatrix}
			0,-{\omega}/{cq^{\pm}_z},1,0
		\end{bmatrix}.
	\end{flalign}
\end{subequations}	
Here, we distinguish the two degenerate modes by the relative angle of their electric field component with respect to the plane of incidence, where the mode with an electric field normal(parallel) to the plane is referred to as the $s$($p$) eigen mode.

In the case of a thin film with $M$ atomic layers sandwiched between two relatively thick (semi-infinite) materials (vacuum in this case), once the electromagnetic eigenmodes of both materials are found, we solve Eq.~\eqref{Eq.discMaxwell} as follows.	Assuming $I$, $J$ layer indices range between 0 and $(M+1)$, where, 0th and $(M+1)$th layers are in the vacuum adjacent to the top and bottom surfaces, respectively, the wave equation can be solved for fields in the material, $I, J=1...M$, thus leaving only the electric and magnetic fields in the vacuum layers next to the surfaces, $I, J=0, M+1$,

\begin{widetext}
	\begin{subequations}		
		\begin{flalign}
			&(\frac{1}{2}\hat{\mathcal{M}}_{0,0}-i\frac{1}{z_{1}-z_{0}}\hat{F})\begin{bmatrix}
				\mathcal{\vec{E}}_{0}/c\\
				\mathcal{\vec{B}}_{0}
			\end{bmatrix}+\sum_{J=1}^M(\frac{1}{2}\hat{\mathcal{M}}_{1,J}+i\frac{\delta_{1J}}{z_{1}-z_{0}}\hat{F})\begin{bmatrix}
				\mathcal{\vec{E}}_{J}/c\\
				\mathcal{\vec{B}}_{J}
			\end{bmatrix}
			=0,\\
			&\sum_{J=1}^M(\frac{\hat{\mathcal{M}}_{I+1,J}+\hat{\mathcal{M}}_{I,J}}{2}+i\hat{F}\frac{\delta_{I+1J}-\delta_{IJ}}{z_{I+1}-z_{I}})\begin{bmatrix}
				\mathcal{\vec{E}}_{J}/c\\
				\mathcal{\vec{B}}_{J}
			\end{bmatrix}
			=0,\ \ \  I=1...M-1\\
			&\sum_{J=1}^M(\frac{1}{2}\hat{\mathcal{M}}_{M,J}-i\frac{\delta_{MJ}}{z_{M+1}-z_{M}}\hat{F})\begin{bmatrix}
				\mathcal{\vec{E}}_{J}/c\\
				\mathcal{\vec{B}}_{J}
			\end{bmatrix}+(\frac{1}{2}\hat{\mathcal{M}}_{M+1,M+1}+i\frac{1}{z_{M+1}-z_{M}}\hat{F})\begin{bmatrix}
				\mathcal{\vec{E}}_{M+1}/c\\
				\mathcal{\vec{B}}_{M+1}
			\end{bmatrix}
			=0.
		\end{flalign}
	\end{subequations}
	Solving this system of equations yields

	\begin{subequations}
		\begin{flalign}
			&(\frac{1}{2}\hat{\mathcal{M}}_{0,0}-i\frac{1}{z_{1}-z_{0}}\hat{F})\begin{bmatrix}
				\mathcal{\vec{E}}_{0}/c\\
				\mathcal{\vec{B}}_{0}
			\end{bmatrix}+\sum_{J=1}^M(\frac{1}{2}\hat{\mathcal{M}}_{1,J}+i\frac{\delta_{1J}}{z_{1}-z_{0}}\hat{F})\begin{bmatrix}
				\mathcal{\vec{E}}_{J}/c\\
				\mathcal{\vec{B}}_{J}
			\end{bmatrix}
			=0,\\
			&\begin{bmatrix}
				\mathcal{\vec{E}}_{J}/c\\
				\mathcal{\vec{B}}_{J}
			\end{bmatrix}=-\hat{G}_{J,M}(\frac{1}{2}\hat{\mathcal{M}}_{M+1,M+1}+i\frac{1}{z_{M+1}-z_{M}}\hat{F})\begin{bmatrix}
				\mathcal{\vec{E}}_{M+1}/c\\
				\mathcal{\vec{B}}_{M+1},
			\end{bmatrix}
		\end{flalign}
	\end{subequations}
	where we define,
	
	\begin{flalign}			
		[\hat{G}^{-1}]_{M,J}&=\frac{1}{2}\hat{\mathcal{M}}_{M,J}-i\frac{\delta_{MJ}}{z_{M+1}-z_{M}}\hat{F}.
	\end{flalign}
	The transfer matrix is then given by,
	\begin{flalign}
		\hat{T}=(\frac{1}{2}\hat{\mathcal{M}}_{0,0}-i\frac{1}{z_{1}-z_{0}}\hat{F})^{-1}\sum_{J=1}^M(\frac{1}{2}\hat{\mathcal{M}}_{1,J}+i\frac{\delta_{1J}}{z_{1}-z_{0}}\hat{F})\hat{G}_{J,M}(\frac{1}{2}\hat{\mathcal{M}}_{M+1,M+1}+i\frac{1}{z_{M+1}-z_{M}}\hat{F}).
	\end{flalign}
\end{widetext}
Eq.~\eqref{eq:D4b} implies that we can restrict our attention to the in-plane components of the electromagnetic field. In this case, the transfer matrix is given by,

\begin{flalign}\label{Eq.Transfer_xy}
	\hat{T}_{\parallel}^{\alpha\beta}=\hat{T}^{\alpha\beta}-\hat{T}^{\alpha z}\frac{1}{\hat{T}^{zz}}\hat{T}^{z\beta}, \ \ \ \alpha,\beta=x,y
\end{flalign}

Eq.~\eqref{Eq.0N_Waves} can be rewritten as

\begin{flalign}\label{Eq.0N_Waves1}
	\sum_{n=s,p}\hat{T}_{\parallel}[\hat{\varphi}_t]^+_{n}\mathcal{A}_{n,t}
	=\sum_{n=s,p}([\hat{\varphi}_0]^-_{n}\mathcal{A}_{n,r}+[\hat{\varphi}_0]^+_{n}\mathcal{A}_{n,i}),
\end{flalign}
where $\mathcal{A}_{n, i}$, $\mathcal{A}_{n,r}$ and $\mathcal{A}_{n,t}$ correspond to the expansion coefficients representing the wave amplitudes of the incident, reflected, and transmitted lights, respectively. Here, the subscript $t$ stands for the transmitted electromagnetic wave, which corresponds to the electromagnetic wave at $(M+1)$th layer, and the subscripts $i$ and $r$ denote the incident and reflected light, respectively at the $0$-th layer. Solving Eq.~\eqref{Eq.0N_Waves1} for the reflected and transmitted light amplitudes in terms of the incident light, we obtain,

\begin{subequations}
	\begin{flalign}
		\mathcal{A}_{n,r}&=\sum_{m=s,p}[\hat{\varphi}_0^{-1}]^-_{n}\hat{T}_{\parallel}[\hat{\varphi}_t]^+_{m}\mathcal{A}_{m,t},\label{Eq.0N_Waves2a}\\
		\mathcal{A}_{n,i}&=\sum_{m=s,p}[\hat{\varphi}_0^{-1}]^+_{n}\hat{T}_{\parallel}[\hat{\varphi}_t]^+_{m}\mathcal{A}_{m,t}.\label{Eq.0N_Waves2b}
	\end{flalign}
\end{subequations}
The transmission coefficients, defined as $t_{mn} = \mathcal{A}_{m,t} / \mathcal{A}_{n,i}$, are obtained by solving Eq.~\eqref{Eq.0N_Waves2b}. These solutions are then substituted into Eq.~\eqref{Eq.0N_Waves2a} to determine the reflection coefficients, $r_{mn} = \mathcal{A}_{m,r} / \mathcal{A}_{n,i}$, which represent the ratio of the reflected $m$-polarized to the incident $n$-polarized electric field. Here, $m,n = s,p$ denote the polarization states.

\section{Bulk response of the extrinsic electro-optic effect}\label{app:bulkresponse}

In this Appendix, we discuss the origin of the bulk electro-optical response seen in Figs.~\ref{fig:fig9}(e) and (g).  To do so, we show that this bulk electro-optic effect occurs in a Drude model, to linear order in the ac field wave vector $q$. The linear-in-$q$ response is applicable to the site-antisymmetric response of our real-space calculation.

We consider a free electron gas (density $n_e$, mass $m$, charge $-e$) within the relaxation‐time (Drude) approximation, driven by a \emph{uniform} dc field $\vec{E}_0$ and a weak ac plane wave
$\vec{E}(\vec{r},t)=\Re\!\left[\vec{E}(\vec{q},\omega)e^{i(\vec{q}\cdot\vec{r}-\omega t)}\right]$ with accompanying magnetic field
$\vec{B}(\vec{q},\omega)$. We keep only terms linear in $\vec{E}_0$ and the ac fields ({\it i.e.}, $\mathcal O(E_0 E_\omega)$). For the electro-optic effect, we will then focus on terms that are bilinear in $\vec{E}_0$ and the ac electric or magnetic field.

Let $\vec{v}=\vec{v}_0+\vec{v}_1(\vec{q},\omega)$ and $n=n_e+n_1(\vec{q},\omega)$ with $|\vec{v}_1|,|n_1|\ll 1$.
The Drude momentum and continuity equations are
\begin{align}
	m\!\left(\partial_t+1/\tau\right)\!\vec{v}
	&= -e\!\left[\vec{E} + \vec{v}\times \vec{B}\right],\\
	\partial_t n + \,\vec{\nabla}\!\cdot\!(n\vec{v})&=0,
\end{align}
where $\tau=\hbar/\eta$ is the relaxation time. The dc drift is
\begin{align}
	\vec{v}_0=-\frac{e\tau}{m}\,\vec{E}_0.
\end{align}
For the ac perturbations (phasor convention $e^{-i\omega t}$), define
\(
D(\omega)\equiv 1/\tau-i\omega.
\)
Linearizing to first order in the ac fields, $\vec{q}$ and $\vec{v}_0$,
\begin{align}
	&m D(\omega)\,\vec{v}_1(\vec{q},\omega)
	= -e\!\left[\vec{E}(\vec{q},\omega)+\vec{v}_0\times\vec{B}(\vec{q},\omega)\right], \label{eq:lin-mom}\\
	&-i\omega\,n_1(\vec{q},\omega) + i n_e\,\vec{q}\!\cdot\!\vec{v}_1(\vec{q},\omega)=0
	\Rightarrow
	n_1=\frac{n_e}{\omega}\,\vec{q}\!\cdot\!\vec{v}_1. \label{eq:lin-cont}
\end{align}

The total current is $\,\vec{J}=-e\,n\,\vec{v}\,$. Keeping cross terms up to linear in $\vec{E}_0$ and the ac fields,
\begin{align}
	\vec{J}(\vec{q},\omega)
	&\simeq -e\,n_e\,\vec{v}_1 - e\,n_1\,\vec{v}_0,\nonumber\\
	&=\frac{e^2n_e}{m(1/\tau-i\omega)}\!\left[\vec{E}(\vec{q},\omega)+\vec{v}_0\times\vec{B}(\vec{q},\omega)\right] - e\frac{n_e}{\omega}(\vec{q}\!\cdot\!\vec{v}_1)\vec{v}_0,\nonumber\\
	&\simeq\frac{e^2n_e}{m(1/\tau-i\omega)}\!\left[\vec{E}(\vec{q},\omega)-\frac{e\tau}{m}\vec{E}_0\times\vec{B}(\vec{q},\omega)\right]\nonumber\\
	&-\frac{e^3n_e\tau}{m^2\omega(1/\tau-i\omega)}(\vec{q}\!\cdot\!\vec{E}(\vec{q},\omega))\vec{E}_0,\nonumber\\
	&\simeq\frac{e^2n_e}{m(1/\tau-i\omega)}\vec{E}(\vec{q},\omega)\\
	&-\frac{e^3n_e\tau}{m^2(1/\tau-i\omega)}\left(\frac{\vec{q}\!\cdot\!\vec{E}(\vec{q},\omega)}{\omega}\vec{E}_0+\vec{E}_0\times\vec{B}(\vec{q},\omega)\right).\nonumber\\
	{J}^{\alpha}(\vec{q},\omega)&
	=\frac{e^2n_e}{m(1/\tau-i\omega)}\left(1+\frac{e\tau}{m\omega}\vec{q}\cdot\vec{E}_0\right){E}^{\alpha}(\omega)\label{eq:eq_F7}\\
	&-\frac{e^3n_e\tau}{m^2(1/\tau-i\omega)}\left(\frac{\vec{q}\cdot\vec{E}(\omega){E}^{\alpha}_{0}+q_{\alpha}\vec{E}_{0}\cdot\vec{E}(\omega)}{\omega}\right),\nonumber
\end{align}

where we used the Maxwell-Faraday equation, \mbox{$\vec{B}(\vec{q},\omega)=\vec{q}\times\vec{E}(\omega)/\omega$}, to convert the ac magnetic field into an ac electric field. 
Equating this expression with \mbox{$J^{\alpha}(\vec{q})=\chi_{\alpha\beta}^{\gamma}(\vec{q}){E}_{0}^{\gamma}{E}^{\beta}(\omega)$}, within Drude approximation, for the nonzero off-diagonal ($\alpha\neq\beta$) elements of the electro-optic response, we obtain, 
\begin{align}
	\frac{\partial\chi_{\alpha\beta}^{\alpha}}{\partial{q}^{\beta}}=\frac{\partial\chi_{\beta\alpha}^{\alpha}}{\partial{q}^{\beta}}=-\tau\frac{e^3n_e}{m^2(1/\tau-i\omega)\omega},\label{Eq.EqS46}
\end{align}
Eq.~\eqref{Eq.EqS46} suggests that the dispersive electro-optic response diverges in the dc limit, $\omega\to0$, although enforcing the optical dispersion, $\omega=c|\vec{q}|$, renders it finite. The linear in $\vec{q}$ dependence results in an electro-optic response, $\chi_{I\alpha,I\beta}^{\gamma}$, that is an odd function of $\vec{r}_{IJ}=\vec{r}_{I}-\vec{r}_{J}$, consistent with the result shown in Fig.~\ref{fig:fig9}(e,g) in the main text, where the $\vec{q}$-dependence can be obtained by using Fourier transformation, $\chi_{\alpha\beta}^{\gamma}(\vec{q})=\sum_{\vec{r}_{IJ}}\chi_{I\alpha,J\beta}^{\gamma}e^{i\vec{r}_{IJ}\cdot\vec{q}}$. The $\tau=\hbar/2\eta$ proportionality of ${\partial\chi_{\alpha\beta}^{\alpha}}/{\partial{q}^{\beta}}$ (i.e., only extrinsic contribution) and also its symmetric tensor form, $\partial\chi_{xz}^{x}/\partial q_z=\partial\chi_{zx}^{z}/\partial q_z$ are also consistent with our numerical calculations. 

We next connect this result to the numerical calculations presented in this work. The direction of the spatially varying ac current and fields in our calculation is fixed along $z$.  This corresponds to the $\vec{q}$ vector of the Drude model. The direction of the dc field in our calculations is fixed along $x$. Therefore, the term proportional to $\vec{q}\cdot\vec{E}_0$ in Eq.~\eqref{eq:eq_F7}—which affects only the longitudinal optical conductivity and does not generate Kerr rotation—is omitted from our results.  The bulk component of $\chi^x_{xz}$ in our results is derived from the term proportional to $(\vec{q}\cdot\vec{E}(\omega))\vec{E}_0$ in Eq.~\eqref{eq:eq_F7}, while the bulk component of $\chi^x_{zx}$ is derived from the term proportional to $(\vec{E_0}\cdot\vec{E}(\omega))\vec{q}$ in Eq.~\eqref{eq:eq_F7}.

\begin{figure}%
	{\includegraphics[scale=0.32,angle=0,trim={0.0cm 0.0cm 0.0cm 0.0cm},clip,width=0.5\textwidth]{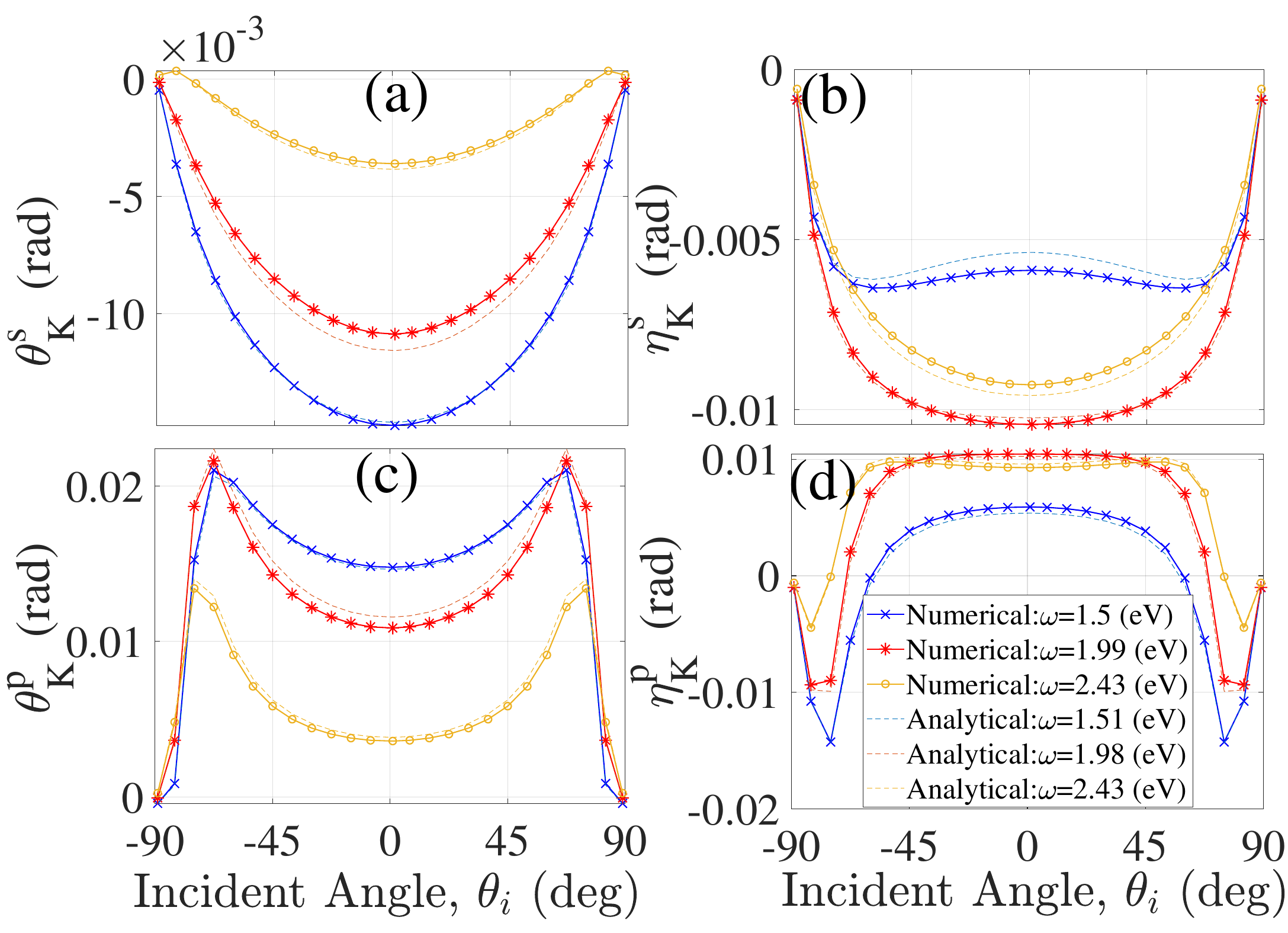}}%
	\caption{Complex Kerr rotation for a 25 monolayer Fe film versus incident angle for (a,b)$s-$ and (c,d)$p-$polarized light calculated using numerical (solid lines) and analytical(dashed lines) at $\hbar\omega=$1.5~eV, 1.99~eV and 2.43~eV.}%
	\label{fig:fig3}%
\end{figure}

\section{Magneto-optical Kerr Rotation in Equilibrium}\label{app:equilibriumFM}

In order to validate the numerical methodology presented in the previous sections, we consider ferromagnetic slabs of bcc Fe(001) and hcp Co(0001) and calculate the corresponding complex Kerr rotation in a \emph{polar} MOKE setup ({\it i.e.,} magnetization along $z$-axis).  The Kerr response in ferromagnetic materials in equilibrium can be cast in terms of a magnetisation–dependent
permittivity tensor
\(\varepsilon_{\alpha\beta}(\vec{m})\) that must satisfy the symmetry
operations of the \emph{magnetic} point group of the crystal
(Neumann’s principle) \cite{Nye1985,birss1964,OPPENEER2001_book}.
Up to second order in the unit magnetization,
\(\vec{m}\), one can write \cite{Fan2016,Zvezdin1997,Visnovsky1986}
\begin{equation}
	\label{eq:eps_expansion}
	\varepsilon_{\alpha\beta}(\vec{m}) =
	\varepsilon^{(0)}_{\alpha\beta}
	+ K_{\alpha\beta\gamma}m_\gamma
	+ G_{\alpha\beta\gamma\delta}m_\gamma m_\delta,
\end{equation}
where \(K_{\alpha\beta\gamma}\) (\(G_{\alpha\beta\gamma\delta}\))
is a third- (fourth-) rank \emph{axial} tensor that is antisymmetric
(symmetric) in the first two indices,  
\mbox{\(K_{\alpha\beta\gamma}=-K_{\beta\alpha\gamma}\)}~ (\mbox{\(G_{\alpha\beta\gamma\delta}=G_{\beta\alpha\gamma\delta}\)}).

For a crystal structure with cubic symmetry, Eq.~\eqref{eq:eps_expansion} reduces to,
\begin{widetext}
	\begin{flalign}
		\hat{\epsilon}^r=
		\begin{pmatrix}
			\epsilon^r_0 & iQm_z & -iQm_y\\
			-iQm_z & \epsilon^r_0 & iQm_x\\
			iQm_y        & -iQ\,m_x        & \epsilon^r_0
		\end{pmatrix}
		+
		\begin{pmatrix}
			B_{1} m_x^2 & B_{2}m_{x}m_y & B_{2}m_xm_z\\
			B_{2}m_xm_y & B_{1} m_y^2 & B_{2}m_ym_z\\
			B_{2}m_xm_z & B_{2}m_ym_z & B_{1} m_z^2
		\end{pmatrix}.
	\end{flalign}    
\end{widetext}
The linear-in-\(\vec{m}\) off-diagonal term
(\(Q\)) is, to the lowest order linear in spin-orbit coupling and odd under time-reversal symmetry, and produces the linear Kerr effects \cite{Dehesa2001}, whereas the quadratic part ($B_{1,2}$) is second order in spin-orbit coupling, even under time-reversal and leads to the quadratic MOKE~\cite{Buchmeier2009}.

\begin{figure}%
	{\includegraphics[scale=0.32,angle=0,trim={0.0cm 0.0cm 0.0cm 0.0cm},clip,width=0.5\textwidth]{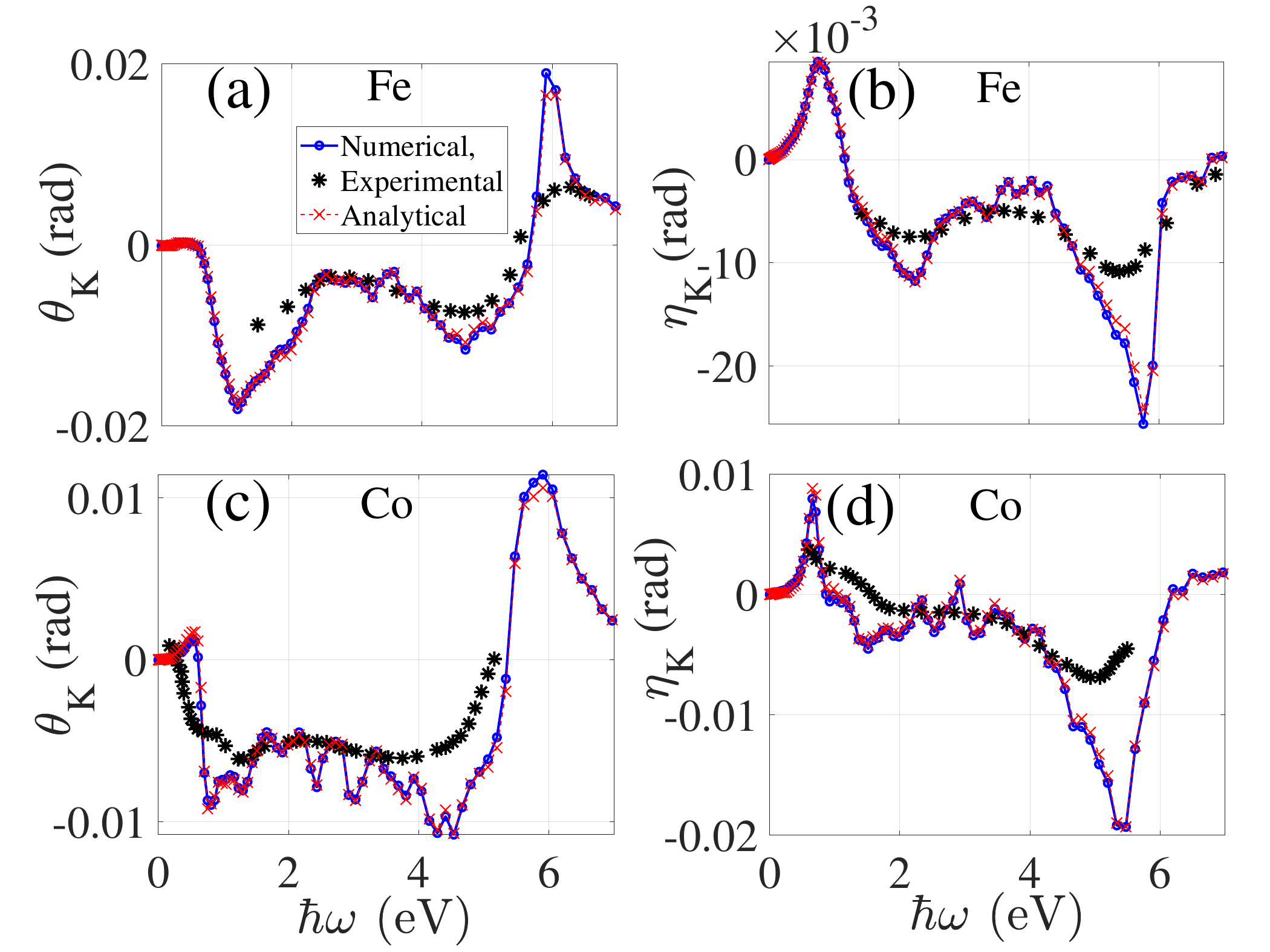}}%
	\caption{Comparison of polar complex Kerr rotation ($\theta_i = 90^{\circ}$) between experimental data (black stars), numerical simulations (blue lines with open circles), and analytical calculations (red lines with stars) for (a,b) Fe and (c,d) Co films. The experimental data are taken from Ref.~\cite{OPPENEER2001_book}.}%
	\label{fig:fig2}%
\end{figure}

Fig.~\ref{fig:fig3} shows the dependence of the Kerr rotation on the incident angle of the electromagnetic wave from the Fe film for three different frequencies, including $\hbar\omega=$1.5~eV, 1.99~eV, and 2.4~eV. The dashed lines correspond to the results obtained using the analytical expression in Eqs.~\eqref{Eq.analytical}. Overall, there is good agreement between the numerically calculated complex Kerr rotation and the analytical results across all incident angles for both $s$- and $p$-polarized light. Notably, at incidence angles less than 45$^\circ$ relative to the normal axis, the Kerr angle for $s$-polarized light tends to decrease as $\theta_i$ increases, while $p$-polarized light exhibits a larger Kerr angle. This behavior can be attributed to the pseudo-Brewster angle, where the reflectivity of $p$-polarized light reaches a non-zero minimum.

Fig.~\ref{fig:fig2} presents the results for the normal incident case versus optical frequency for Fe and Co films, where we also included experimental data as black star symbols. For comparison, the results using the analytical expression for the Kerr rotation for the semi-infinite ferromagnets are also included as red lines with cross symbols. The bulk dielectric tensor in the analytical approach was calculated by summing over the layer indices in the two-point dielectric tensor, divided by the total number of layers. The numerical results are calculated by constructing a superlattice that consists of a sufficiently large number of slabs, each consisting of 25 monolayers. The results demonstrate an excellent agreement between analytical and numerical methods and a good agreement with the experimental measurements. The deviation between the experiment and theory can be attributed to the smearing effect due to the other collective excitations in the material and the resulting shorter relaxation time at high optical frequencies.

{\it Angular dependence of field-induced Kerr Rotation:}
\begin{figure}%
	{\includegraphics[scale=0.32,angle=0,trim={0.0cm 0.0cm 0.0cm 0.0cm},clip,width=0.5\textwidth]{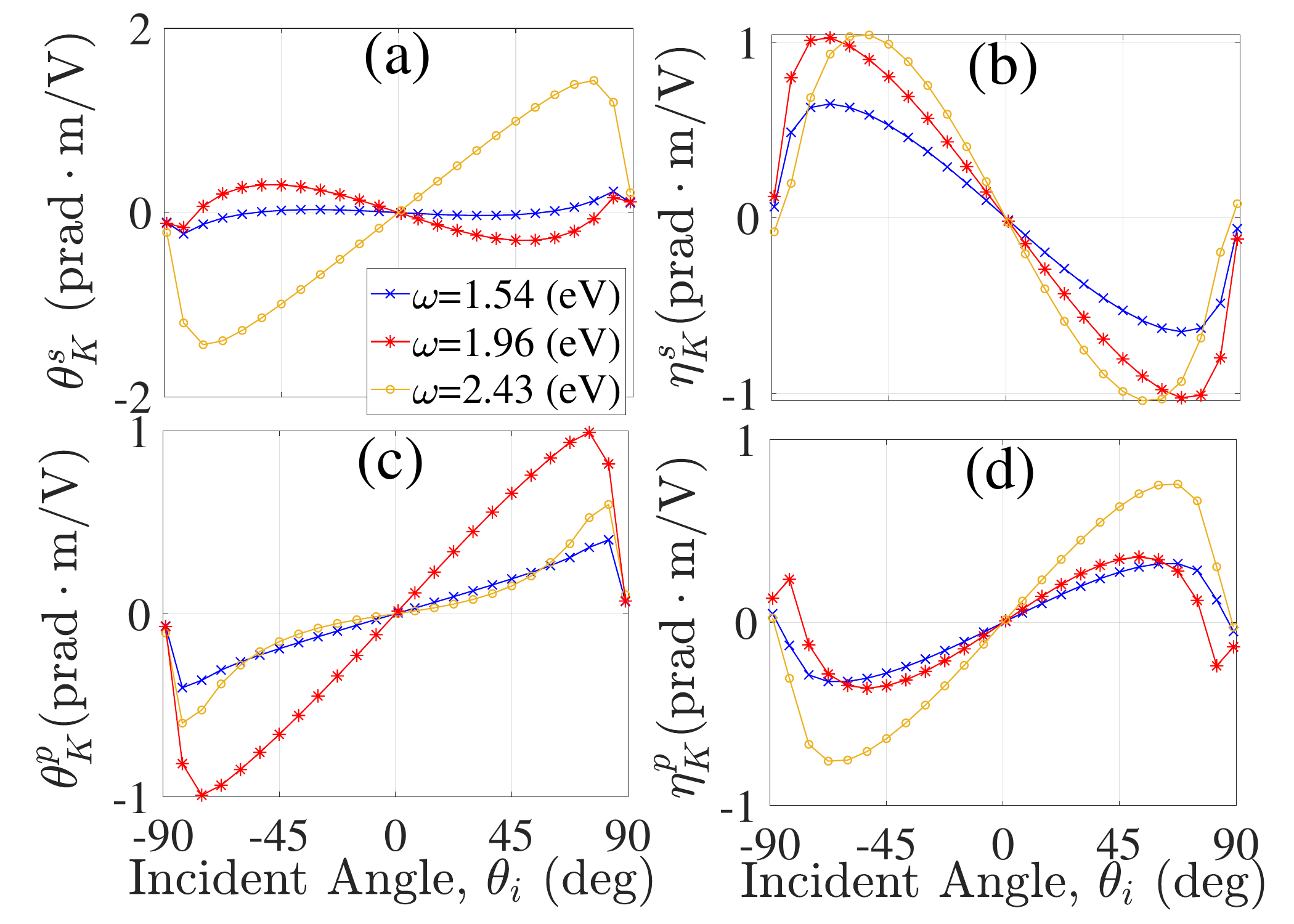}}%
	\caption{Electric field induced complex Kerr rotation for 55 monolayers Pt film versus incident angle for (a,b)$s-$ and (c,d)$p-$polarized light calculated for optical frequencies, $\hbar\omega=$1.5~eV, 1.99~eV and 2.43~eV.}%
	\label{fig:fig8_1}%
\end{figure}
In the presence of an in-plane current along the $x$-axis, the induced off-diagonal conductivity tensor has an $xz$ component, which requires an oblique incidence in a \emph{longitudinal} MOKE setup, in order to yield a finite Kerr rotation.  
Fig.~\ref{fig:fig8_1} presents the results for the electric-field induced complex Kerr rotation in 55 monolayers Pt film versus incident angle for three different frequencies, including $\hbar\omega=$1.5~eV, 1.99~eV and 2.43~eV. Consistent with the case of the longitudinal Kerr rotation, we observe an odd dependence of the Kerr rotation versus the incident angle for both light polarizations.

{\it Analytical Expressions:}
The complex Kerr rotations from a material with off-diagonal dielectric tensor elements can be evaluated using~\cite{You1996}
\begin{subequations}	\label{Eq.analytical}	
	\begin{flalign}
		r_{ss}&=\frac{\cos(\theta_i)-n\cos(\theta_t)}{\cos(\theta_i)+n\cos(\theta_t)},\\
		r_{pp}&=\frac{n\cos(\theta_i)-\cos(\theta_t)}{n\cos(\theta_i)+\cos(\theta_t)},\\
		r_{sp}&=\frac{\cos(\theta_i)(\tan(\theta_t)\epsilon^r_{xz}-\epsilon^r_{xy})}{n(n\cos(\theta_i)+\cos(\theta_t))(\cos(\theta_i)+n\cos(\theta_t))},\\
		r_{ps}&=\frac{\cos(\theta_i)(\tan(\theta_t)\epsilon^r_{zx}+\epsilon^r_{yx})}{n(n\cos(\theta_i)+\cos(\theta_t))(\cos(\theta_i)+n\cos(\theta_t))},
	\end{flalign}
\end{subequations}
where, $n=\sqrt{\epsilon^r_{xx}}$ and $n\sin(\theta_t)=\sin(\theta_i)$. It should be noted that equations in \eqref{Eq.analytical} are derived perturbatively, retaining terms only up to linear order in the off-diagonal elements of the dielectric tensor. In both the analytic derivation and the numerical code we adopt a fixed polarization basis: the $s$-polarized unit vector $\vec{e}_{s}$ lies in the film plane along the laboratory $x$-axis, while the $p$-polarized unit vector $\vec{e}_{p}$ is chosen so that $\vec{e}_s\times\vec{e}_{p}=\vec{q}/q$. Because the $z$-component of the propagation vector, $\vec{q}$, reverses for the reflected beam, the in-plane $y$-component of $\vec{e}_{p}$ flips sign, whereas $\vec{e}_{s}$ remains unchanged; this convention preserves a right-handed coordinate system for both incident and reflected light and an opposite sign for $r_{pp}$ compared to $r_{ss}$ at normal incidence. This sign reversal makes the polar MOKE response for $p$-polarized incidence opposite in sign to that for $s$-polarized incidence, as seen in Fig.~\ref{fig:fig3}.

\section{Finite Size Extrapolation method}\label{app:finite_size_extra}

\begin{figure}
	\centering
	\includegraphics[scale=0.32,angle=0,trim={1.0cm 0.0cm 0.0cm 0.0cm},clip,width=0.5\textwidth]{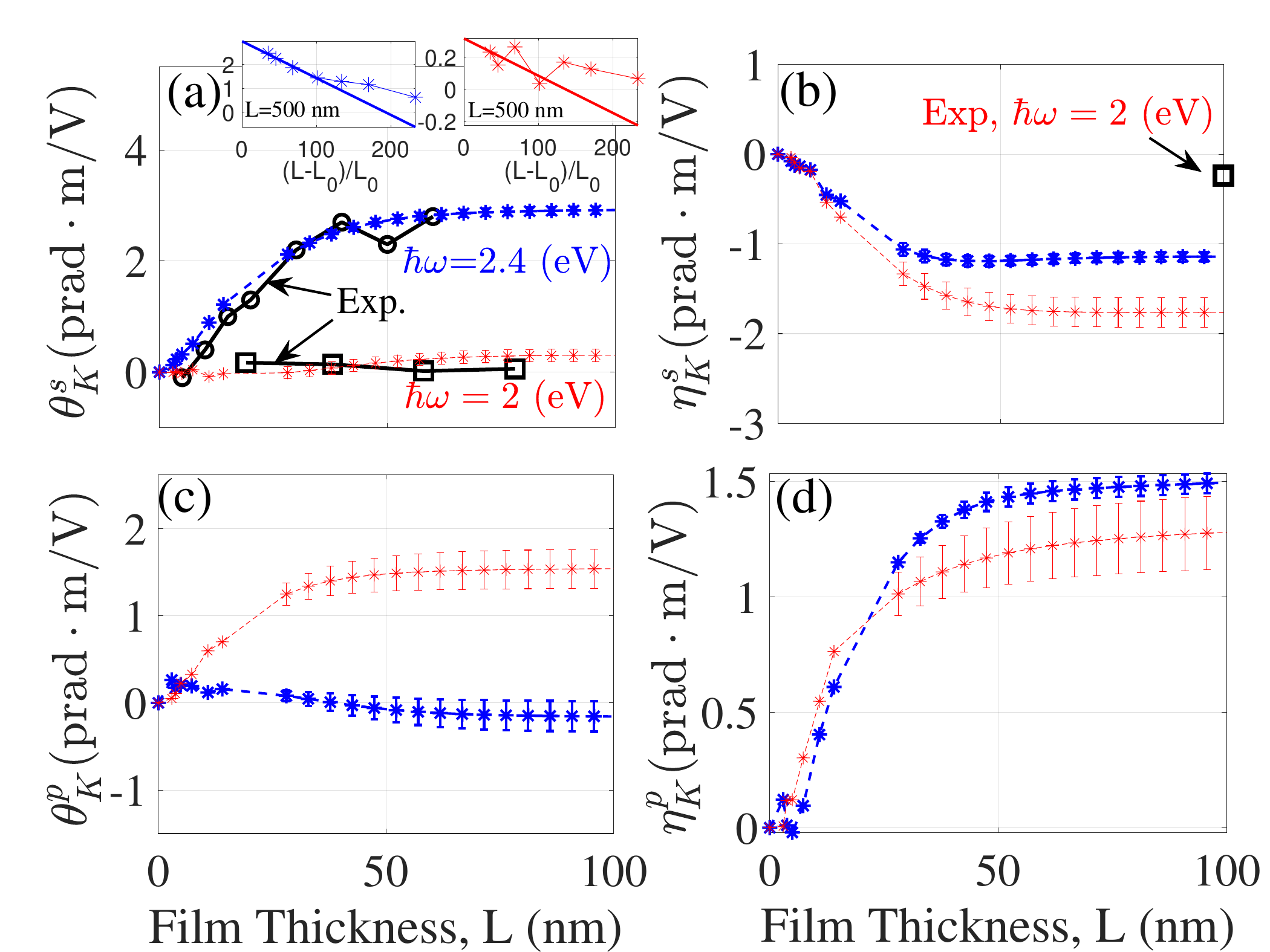}%
	\caption{Total Kerr angle (a) and ellipticity (b) for an $s$-polarized incident light versus Pt film thickness for $\hbar\omega=2$~eV and $\hbar\omega=2.4$~eV, shown as red and blue lines with star symbols, respectively. The insets shows the Kerr angle versus $(L-L_0)/L_0$ for the case of a fixed total thickness $L=500$ nm, and varying unit cell thicknesses $L_0$. In the extrapolation scheme, we use only the first four points corresponding to the thickest unit cell film thicknesses, $L_0$. (c) and (d) presents the corresponding results for $p$-polarized incident light. The thick black lines are from experimental measurements, as reported in Ref.~\cite{Stamm2017} and Ref.~\cite{Marui2024}. The first few points without errorbar are calculated using a single Pt film with a substrate with $n_{\rm sub}=4$, and the rest are from the finite size extrapolation method. An energy broadening value of $\eta=25$ meV was chosen in the numerical electronic calculations. The incidence angle in all cases is 45$^{\circ}$. The error-bar amplitudes denote the 95th-percentile ranges from the fit.}%
	\label{fig:fig12F}%
\end{figure}

As an alternative to the truncation scheme used in the main text, we compute the full electro-optic response throughout a Pt superlattice of total thickness $L$, composed of films of thickness $L_0$, and then extrapolate by regressing the results versus \((L-L_0)/L_0>1\) at fixed total film thickness, $L$. For sufficiently thick films, $L_0$, where finite-size corrections are small, the complex Kerr response (rotation and ellipticity) admits the asymptotic form,
\begin{equation}
	\theta^{s,p}_K(L,L_0)=\theta_K^{s,p}(L,L)+C^{s,p}(L)\frac{L-L_0}{L_0}+\mathcal{O}\!\left(\frac{(L-L_0)^2}{L_0^2}\right), 
	\label{eq:1overL}
\end{equation}
and analogously for $\eta^{s,p}_K$. 
Since, $(L-L_0)/L_0$ corresponds to the number of ``interfaces" residing in the bulk of the optical superlattice setup, a linear regression of the complex \(\theta_K\) against \((L-L_0)/L_0\) therefore yields the intercepts \(\theta_K^{s,p}(L,L)\), that correspond to the case without the interfaces in the bulk.

Fig.~\ref{fig:fig12F} shows the results of the finite size extrapolation versus total thickness of the Pt super-lattice structure for photon energies \(\hbar\omega=2.0~\mathrm{eV}\) and \(2.4~\mathrm{eV}\). The first few points in Fig.~\ref{fig:fig12F} (without error bars) come from single-film calculations on a substrate with refractive index \(n_{\mathrm{sub}}=4\); the remaining points are obtained through the finite-size extrapolation workflow implied by Eq.~\eqref{eq:1overL}. The insets of panel (a) show representative \(\theta_K\) versus \((L-L_0)/L_0\) trends and their linear fits for $\hbar\omega=2.4$ eV (left inset) and $\hbar\omega=2$ eV (right inset). Error bars represent 95 $\%$ confidence intervals, calculated from the standard deviation of the residuals of the linear least-squares fit.

Panels (a) and (b) report the total Kerr rotation and ellipticity for \(s\)-polarized incidence, while panels (c) and (d) show the corresponding quantities for \(p\)-polarized incidence. Blue symbols denote \(\hbar\omega=2.4~\mathrm{eV}\) and red symbols denote \(\hbar\omega=2.0~\mathrm{eV}\). Experimental benchmarks (thick black lines) are reproduced from Refs.~\cite{Stamm2017,Marui2024}. 

To ensure asymptotic validity of Eq.~\eqref{eq:1overL}, the fits are restricted to higher thicknesses for which higher-order terms are negligible and the residuals versus \(1/L_0\) are structureless.


\bibliographystyle{apsrev4-2}
\bibliography{ref}{}

\end{document}